\documentclass{aa}

\usepackage[varg]{txfonts}

\usepackage{multirow}
\usepackage{graphicx}

\def\ms{\mbox{$M_{\rm s}$}}
\def\mh{\mbox{$M_{\rm h}$}}   
\def\mstar {h^{-2}~\rm M_{\sun}}
\def\mhalo {h^{-1}~\rm M_{\sun}}
\def\msun {\rm M_{\sun}}

\bibpunct[; ]{(}{)}{;}{a}{}{;}

\begin{document}

\title{Isolated elliptical galaxies in the local Universe}
%\subtitle{Isolated ellipticals}

\author{I.~Lacerna\inst{\ref{ia-puc},\ref{aiuc},\ref{mpia}} \and H.~M.~Hern\'andez-Toledo\inst{\ref{unam}} 
\and V.~Avila-Reese\inst{\ref{unam}} \and J.~Abonza-Sane\inst{\ref{unam}} \and 
A.~del Olmo\inst{\ref{iaa}}  
}  

\institute{Instituto de Astrof\'isica, Pontificia Universidad Cat\'olica de Chile, Av. V.~Mackenna 4860, Santiago, Chile \\
\email{ialacern@astro.puc.cl} \label{ia-puc} 
\and 
Centro de Astro-Ingenier\'{\i}a, Pontificia Universidad Cat\'olica de Chile, Av. V.~Mackenna 4860, Santiago, Chile\label{aiuc} 
\and 
Max Planck Institute for Astronomy, K\"onigstuhl 17, D-69117 Heidelberg, Germany \label{mpia} 
\and 
Instituto de Astronom\'{\i}a, Universidad Nacional Aut\'onoma de M\'exico, A.P. 70-264, 04510 M\'exico D. F., M\'exico \label{unam} 
\and
Instituto de Astrof\'isica de Andaluc\'ia IAA – CSIC, Glorieta de la Astronom\'ia s/n, 18008 Granada, Spain \label{iaa}
}
%A. del Olmo \altaffilmark{3}}
%\affil{Instituto de Astrofisica de Andalucia.  Seoul 130-012, Republic of Korea}

%\altaffiltext{1}{E-mail: hector@astro.unam.mx}
%\altaffiltext{1}{E-mail: jvazquez@astroscu.unam.mx}
%\altaffiltext{2}{E-mail: chony@iaa.es}
%\altaffiltext{2}{E-mail: cbp@kias.re.kr}

\abstract
%Context
{We have studied a sample of 89 very isolated, elliptical galaxies
%at $z$ = 0.037 on average
at $z < 0.08$  
and compared their properties with
elliptical galaxies located in a high-density environment such as
the Coma supercluster. %, where is more common to find these galaxies. 
}
%Aims
{Our aim is to probe the role of  environment on
the morphological transformation and quenching of elliptical galaxies
as a function of mass. In addition, we %aim to 
elucidate the nature of a particular set of blue and star-forming isolated ellipticals identified here.
}
%Methods
{We %use photometric and spectroscopic properties based on SDSS data to 
studied physical properties of ellipticals, such as color, specific star formation rate, galaxy size, and stellar age, as a function of stellar mass and environment based on SDSS data. We analyzed  the blue and star-forming isolated ellipticals in more detail, through photometric characterization 
using GALFIT, and infer their star formation history %. along with a population synthesis model 
using STARLIGHT.
}
%Results
{
%All elliptical galaxies more massive than 10$^{11}$ $\msun$ are passive. As the mass is smaller, both isolated and Coma 
%ellipticals become bluer though they remain in the red sequence on average. 
Among the isolated ellipticals $\approx 20\%$ are blue, $\lesssim 8\%$ are star forming, 
and $\approx 10\%$ are recently quenched, 
while among the Coma ellipticals $\approx 8\%$ are blue and just $\lesssim 1\%$ are star forming or recently quenched.
There are four isolated galaxies ($\approx 4.5\%$) that are blue and star forming at 
the same time. %, whereas there are not galaxies with this feature in Coma. 
These galaxies, %blue star-forming isolated ellipticals, 
with masses between $7 \times 10^{9}$ and  $2 \times 10^{10}$ $\mstar$, are also the youngest galaxies with light-weighted stellar ages $\lesssim 1$ Gyr and exhibit bluer colors toward the galaxy center.  %, which could indicate recent processes of star formation for them. 
Around 30 -- 60\% of their present-day luminosity, but only $< 5\%$ of their present-day mass, is due to star formation in the last 1 Gyr.  
%On the other hand, most of the elliptical galaxies with emission lines are AGNs. However, the fraction of those classified as a mixture of emission coming from a circumnuclear star-forming region and central low-luminosity AGN in the BPT diagram is larger for the isolated ellipticals than for those in the Coma cluster (12\% and 6\%, respectively).
}
%Conclusions
{
The processes of morphological transformation and quenching
seem to be in general independent of environment since most of elliptical galaxies are 
`red and dead', %\LEt{According to AA style, please avoid the use of quotation marks unless you are actually quoting something. Please check for this throughout the paper.} 
%\LEm{We replace quotation marks for apostrophes in some cases. In other cases, we avoid the use of quotation marks unless we are actually quoting something.} 
although the transition to the red sequence should be faster for isolated ellipticals. 
%than for cluster ones.
In some cases, the isolated environment seems to propitiate the rejuvenation of ellipticals by recent ($< 1$ Gyr) cold gas accretion. 
}


\maketitle

\section{Introduction} 
\label{S1}

An essential aspect in the theory of formation and evolution of galaxies is the understanding of the mechanisms behind their 
morphological transformation and quenching of star formation. Among the wide diversity of morphological types, elliptical (E) 
galaxies \citep[so defined by][]{Hubble1926} seem to be those in the final stage of transformation to quiescent objects 
with regular and smooth structures \citep[cf.][]{Vulcani+2015}.  As a result of their spheroidal and compact structure, supported against gravity
by velocity dispersion, E galaxies have been proposed to be the result of major and minor mergers of disk galaxies, after which the gas
would have been depleted and star formation quenched. %\citep[][]{}. 
The mergers could have happened early, when the disks were gaseous, 
or late, between gas-poor stellar galaxies (dry mergers). Moreover, in general nearby E galaxies are located in high-density 
environments \citep[][]{Oemler1974, Dressler1980, Bamford+2009}, they have red colors and low values of specific star formation 
rates (i.e., they are quiescent), and their stellar population are mostly old with high metallicities and $\alpha/$Fe ratios 
\citep[e.g.,][for a review and more references see \citealp{Blanton+2009}]{RobertsHaynes1994,Kauffmann1996, 
Kuntschner2000,Baldry+2004,Bell+2004,Thomas+2005,Kuntschner+2010}. 
% Algo de Renzini ?!.
All these facts motivated the idea that the %``red, dead, and compact''
`red and dead'
%IVAN: omito "compact" porque hay una literatura reciente sobre compact Es (algunas aisladas), pero son galaxias mucho mas pequeñas (r_eff <= 0.5 kpc) que no tienen relacion con las classic Es de las que estamos hablando.
galaxies formed long ago by violent processes in a 
high-density environment that contributed to avoid further gas accretion into the galaxy. 

Recent detailed observational studies have shown that ellipticals, in spite of the fact that they are the most regular galaxies, present
more complex structures and more variations in their properties than previously thought \citep[see, e.g.,][]{Blanton+2009}. 
A relevant question is how the properties of E galaxies vary with environment. Since the prevailing mechanism
of E galaxy formation is that of major mergers 
\citep[e.g.,][]{Hernquist1993,Kauffmann1996,Tutukov+2007,Schawinski+2014},
%--
which  commonly happen in dense regions before they virialize, %--, 
the environment is then expected to play a key role in the properties of ellipticals and  their formation 
histories and quenching processes. However, it could be that local galaxy-halo processes and the mass scale rather 
than external processes are mostly responsible for the quenching and general properties of the post-merger systems. 
Hence, the study of isolated E galaxies is important since, in this case, the quenching mechanisms associated with the 
group/cluster environment (starvation, tidal,  and ram-pressure gas stripping, etc.) are not acting.  In general, isolated 
galaxies are optimal objects for constraining the internal physical processes that drive galaxy evolution. 

The question of formation of E galaxies in isolated environments is on its own of great interest. Are these galaxies, for 
a given mass, different from those in clusters? Do both populations follow the same correlations with mass? 
Semianalytic models in the context of the popular $\Lambda$ cold dark matter ($\Lambda$CDM) 
cosmology predict that on average the ellipticals formed in field haloes should have stellar ages comparable to those 
formed in rich clusters. However, in the field environment a more significant fraction of ellipticals with younger 
stellar populations is predicted than in clusters \citep{Kauffmann1996, Niemi+2010}. This may indicate different formation histories.
Theoretical models suggest that ellipticals in clusters form through dissipative infall of gas and numerous mergers
that took place at early epochs (5 to 10 Gyr ago), whereas some field ellipticals form through recent major mergers and 
are still in the process of accreting cold gas. 
%These galaxies are classified as ellipticals simply because they may have not yet had the time to accrete a new disk following their last major merger event.

In most of the previous observational works, early-type (E and S0/a) galaxies in general were studied.
These works find that early-type galaxies are mostly red/passive, but there is also a fraction of blue/star-forming objects;  
this blue fraction increases as  the mass is smaller and the environment is less dense  \citep[e.g.,][]{Schawinski+2009,Kannappan+2009,
Thomas+2010,McIntosh+2014,Schawinski+2014,Vulcani+2015}. The fraction of blue early-type galaxies at masses larger than 
log$(\ms/ \msun)\approx 11$ is virtually null, showing that the most massive galaxies formed very early and 
efficiently quenched  their growth by star formation. Interesting enough, the trends seen at $z\sim 0$ are similar at higher redshifts, although 
the fractions of blue early-type galaxies increase significantly with $z$ \citep{Huertas-Company+2010}. For E galaxies in the field, 
which may include galaxies in loose and poor groups with a dynamical masses $< 10^{13}$ $\msun$,  it has been found 
that their colors are bluer on average and show more scatter in color than ellipticals in rich groups or clusters 
\citep{deCarvalho-Djorgovski1992}. Regarding pure E galaxies in very isolated environments, the samples in these studies are usually composed of only a few bright (massive) objects
%there are only a few 
%observational studies in the literature 
\citep[e.g.,][]{Colbert+2001,Marcum+2004,Reda+2004,Denicolo+2005,Collobert+2006,HauForbes2006,Smith+2010,Lane+2013,Lane+2015,Richtler+2015,Salinas+2015}.
In contrast to these works, \cite{Smith+2004} and \cite{Stocke+2004} presented  relatively large samples of 32 and 65 isolated E galaxies, respectively, but they did not consider the radial velocity separation of companion galaxies to classify an isolated candidate.

In view of the shortage of observational samples of well-defined E (pure spheroidal) galaxies in extreme isolation, we
present here a relatively complete sample of these galaxies in a large mass range and compare some of their
properties to those ellipticals located in a high-density environment, the Coma supercluster.
Our sample comes from the catalog of local isolated galaxies by \citet[][]{Hernandez-Toledo+2010}, which includes
the redshift information creating a robust isolated sample of pure Es.
We explore whether very isolated ellipticals differ in some photometric, spectroscopic, and structural properties
with those of the Coma supercluster, and whether both isolated and high-density ellipticals follow similar correlations with mass. 
Our final aim is to probe the role of environment on the morphological transformation and quenching 
of E galaxies as a function of mass. 
Isolated ellipticals, whose results are very different 
from those in the cluster environment, are studied in more detail.
%We will see that they are only a few, all of them
In particular, we  focus on a set of blue and star-forming 
(hereafter SF) galaxies. 
In case the reason for their blue colors and recent star formation activity is due to a rejuvenation 
process produced by the recent accretion of cold gas, these isolated elliptical galaxies can be used as unique `sensors' of the gas 
cooling from the cosmic web.  
 
The outline of the papers is as follows. The selection criteria of isolated galaxies along with the data set of galaxies in the Coma 
supercluster are described in Sect. \ref{S2}. We present the results with the properties and mass dependences of E galaxies in 
Sect. \ref{secMs}.  The implications of our results are discussed in Sect. \ref{implications}. The photometric and spectroscopic 
analysis of the particular subsample of blue and SF isolated elliptical galaxies is presented in Sect. \ref{blueSF_Es}. Finally, our 
conclusions are given in Sect. \ref{Conclusions}.
 
Throughout this paper we use the reduced Hubble constant $h$, where $H_0 = 100$ $h$ km s$^{-1}$ Mpc$^{-1}$, with the following dependencies: %dependences: 
stellar mass in $\mstar$, absolute magnitude in $+ 5$ log$(h)$, 
size and physical scale in $h^{-1}$ kpc, and 
halo mass in $\mhalo$,  
unless the explicit value of $h$ is specified.

%============================================================
\section{Data and selection criteria} \label{S2}
%============================================================   

Our main goal is to study the properties of local elliptical galaxies in very isolated environments. For this, we use a particular 
galaxy sample described in Sect. \ref{sample-isolated}. In order to compare some of the properties of these galaxies with those in 
a much denser environment, where ellipticals are more frequent, we use a compilation of %these 
elliptical galaxies
%\LEt{Or, please specify what "them" refers to if this is incorrect.} 
in the Coma supercluster as described in Sect. \ref{sample-Coma}.

\subsection{Isolated elliptical galaxies}
\label{sample-isolated}

The isolated elliptical galaxies studied here come from the UNAM-KIAS catalog of \cite{Hernandez-Toledo+2010}; this paper gives more details. Here we briefly refer to the sample and selection criteria.%\LEt{Please ensure that all abbreviations are written out at first mention, followed by the abbreviation in parentheses (even if you have already introduced them in the Abstract). After that please use only the abbreviation. Please check for your use of individual abbreviations throughout the paper. Use the lower case for terms when introducing acronyms. All acronyms need to be introduced for all abbreviations except for units of measurement and the cardinal directions (N, SE, etc.). The acronyms of names of instruments or telescopes can also be introduced when appropriate..} 

The Sloan Digital Sky Survey \citep[SDSS;][]{York+2000,Stoughton+2002} produced two galaxy samples. One is a flux-limited sample to extinction corrected apparent 
Petrosian $r$-band magnitudes of 17.77 (the main galaxy sample), and a color-selected and flux-limited sample extending to $r_{\rm Pet}=19.5$ 
(the luminous red galaxy sample). Galaxies with $r$-band magnitudes in the range $14.5 \leq r_{\rm Pet} < 17.6$ were selected from the DR4plus 
sample that is close to the SDSS Data Release 5 \citep{DR5+2007}. %(Adelman-McCarthy et al. 2007). 
The survey region covers 4464 deg$^2$, containing 312\,338 galaxies. 
\cite{Hernandez-Toledo+2010} attempted to include brighter galaxies, but 
the spectroscopic sample of the SDSS galaxies is not complete for $r_{\rm Pet} < 14.5$. 
Thus, they searched in the literature and 
borrowed redshifts of the bright galaxies without SDSS spectra to increase the spectroscopic completeness. The final data set consists 
of 317\,533 galaxies with known redshift and SDSS photometry. 

The isolation criteria is specified by three parameters. The first is the extinction-corrected Petrosian $r$-band apparent magnitude 
difference between a candidate galaxy and any neighboring galaxy, $\Delta m_r$. The second is the projected separation to the neighbor 
across the line of sight, $\Delta d$. The third is the radial velocity difference, $\Delta V$.  Suppose a galaxy $i$ has a magnitude 
$m_{r,i}$ and $i$-band Petrosian radius $R_i$. It is regarded as isolated with respect to potential perturbers if the separation $\Delta d$ 
between this galaxy and a neighboring galaxy $j$ with magnitude $m_{r,j}$ and radius $R_j$ satisfies the conditions
\begin{equation}
\Delta d \ge 100\times R_j   \\
\label{eq1_UNAM-KIAS}
\end{equation}
\begin{equation}
\textrm{or }  \rm
\Delta V \ge 1000 \textrm{ km s}^{-1}, 
\end{equation}
or  the conditions
\begin{equation}
\Delta d < 100 \times R_j  \\
\end{equation}
\begin{equation}
\Delta V < 1000 \textrm{ km s}^{-1}  \\
\end{equation}
\begin{equation}
m_{r,j} \ge m_{r,i} + \Delta m_r, 
\label{eq5_UNAM-KIAS}
\end{equation}
for all neighboring galaxies. Here $R_{j}$ is the seeing-corrected Petrosian radius of galaxy $j$, measured in $i$-band using elliptical 
annuli to consider flattening or inclination of galaxies \citep{Choi+2007}. 
\cite{Hernandez-Toledo+2010} chose  $\Delta m_r = 2.5$. Using these criteria, 
they found a total of 1548 isolated galaxy candidates. We note that a magnitude difference of 2.5 in
this selection criteria translates into a factor of about 10 in brightness similar to that imposed by \cite{Karachentseva1973}.

In \citet{Hernandez-Toledo+2010}, isolated elliptical galaxies were classified after some basic image 
processing and presented in mosaics of images including surface brightness profiles and the corresponding geometric profiles 
(ellipticity $\epsilon$, Position Angle PA and $A_{4}/B_{4}$ coefficients of the Fourier series expansions of deviations of a pure ellipse) 
from the $r$-band images to provide further evidence of boxy/disky character 
%\LEt{"Boxiness" is a word, but I find no evidence that "diskyness" is a word. You can substitute "character" with another, more appropriate noun, if you like.}
and other structural details. A galaxy was judged to be an 
elliptical if the $A_{4}$ parameter showed: 1) no significant  boxy ($A_{4} < 0$) or disky ($A_{4} > 0$) trend in the outer parts, or 2) a 
generally boxy ($A_{4} < 0$) character in the outer parts. We inspected for the presence/absence of 3) a linear component in the surface brightness-radius diagram. Morphologies were assigned according to a numerical code following the HyperLeda\footnote{http://leda.univ-lyon1.fr/} 
database convention; in particular for early-type galaxies, the following $T$ morphological parameters \citep{Buta+1994} are applied: $-5$ for E, $-3$ for E-S0, $-2$ for S0s, and 0 for S0a types. In the UNAM-KIAS sample there are 250 isolated early-type galaxies that satisfy $T \le 0$ (E/S0), 
where 92 galaxies are ellipticals ($T\le -4$), which is $\approx 6\%$ of the sample.

\subsection{Elliptical galaxies in the Coma supercluster}
\label{sample-Coma}

\begin{figure}
%\resizebox{\hsize}{!}\includegraphics{Hgi_Ms.eps}
\includegraphics[width=9.8cm]{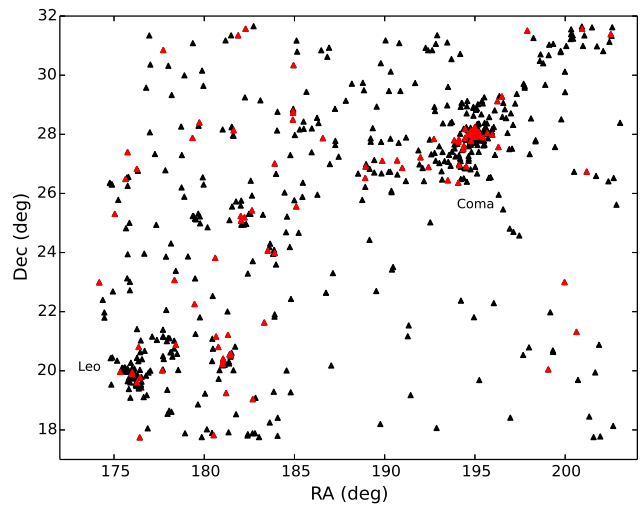}
\caption{
Distribution in right ascension and declination of galaxies in the Coma supercluster with spectroscopic redshifts in the range
$4000 < cz < 9500$ km s$^{-1}$. Red points correspond to elliptical galaxies. Coma and Leo clusters (Abell 1656 and Abell 1367, respectively) are indicated in the plot.
}
\label{Coma_pos}
\end{figure}

To perform a comparative study of the isolated elliptical galaxies in the UNAM-KIAS catalog, we have compiled a sample 
of elliptical galaxies in a dense environment such as the Coma supercluster. 
This region is composed of the Coma and Leo clusters (Abell 1656 and Abell 1367, respectively) along with other galaxies in the filaments that connect these two rich clusters.
To that purpose, we retrieve available data through 
the GOLDMine Database \citep{Gavazzi+2003}. 
From the list of galaxies in the Coma supercluster ($\sim$ 1000 objects), the quoted CGCG (Catalogue of Galaxies and of Clusters of Galaxies) 
principal name was used to cross-correlate with the HyperLeda database. In the latter catalog, 915 galaxies have morphological classification, where 131 objects correspond to pure elliptical galaxies.
There are 113 elliptical galaxies with spectroscopic redshifts from the SDSS database using CasJobs.\footnote{See http://casjobs.sdss.org/CasJobs/}
We select galaxies in the redshift range of $4000 < cz < 9500$ km s$^{-1}$  
%Four galaxies have very low redshifts ($z < 0.005$) to be classified 
as true members of the Coma supercluster \citep[][]{Gavazzi+2014}. With this, we obtain a sample of 102 elliptical galaxies with a mean redshift of $0.023 \pm 0.003$ and $r_{\rm Pet} < 15.4$.
Figure \ref{Coma_pos} shows the spatial distribution of these galaxies in the Coma supercluster.
Although some Es are not members of the two clusters, they probably belong to groups or regions in the outskirts of the clusters, which correspond to higher density environments compared to the low-density environment of isolated galaxies. Indeed, we checked that none of the elliptical galaxies %\LEt{Or, "none of the elliptical galaxies...".} 
in the Coma supercluster is classified as an isolated object by following eqs. (\ref{eq1_UNAM-KIAS})-(\ref{eq5_UNAM-KIAS}). 
In addition, we  checked that the overall results and conclusions in this paper do not change if the velocity range is reduced (e.g.,  $6000 < cz < 8000$ km s$^{-1}$).

%%%Fig1 

\subsection{Physical properties} 

Color measurements for the isolated galaxies and galaxies in the Coma supercluster were taken from the SDSS 
database with extinction corrected $modelMag$ magnitudes  (\verb|dered| parameter in CasJobs). %\footnote{See http://casjobs.sdss.org/CasJobs/}). 
This magnitude is defined as the better of two magnitude fits: a pure de Vaucouleurs profile and a pure exponential profile.  

We transform the SDSS colors to $B - R$ colors 
and the absolute magnitude in the $r$-band %is transformed 
to the $R$-band in
Sect. \ref{secMs}
following the equations, suggested by \cite{Niemi+2010}, written as  
\begin{eqnarray} 
B - R = (g - i) + 0.44 \textrm{} 
\label{eq3_Niemi10}
\end{eqnarray}
%
%Similarly, the absolute magnitude in the $r$-band is transformed to the $R$-band by using
%
\begin{eqnarray} 
M_R = M_r -0.35 \textrm{.} 
\label{eq2_Niemi10}
\end{eqnarray}

We use the following stellar mass-to-light ratio of \cite{Bell+2003} to estimate the stellar mass, \ms, for the galaxy samples: %. Specifically,
\begin{eqnarray} 
\lefteqn{ \textrm{log}(M_s / \mstar)  =  -0.306 + 1.097[^{0.0}(g_{Pet}-r_{Pet})] } \nonumber\\ 
& \qquad & - 0.1 -0.4 (^{0.0}M_{r_{Pet}} - 5 \textrm{ log($h$)} - 4.64) \textrm{,}
\label{logMsBell}
\end{eqnarray}
where $^{0.0}(g_{Pet}-r_{Pet})$ is the color with Petrosian magnitudes (\verb|petroMag| parameter in CasJobs), which are measured 
within a circular aperture defined by the shape of the light profile. In addition to the correction for Galactic extinction \citep{Schlegel+1998}, we perform the
$K$-correction and evolution correction at $z=0$ %via
with \verb|kcorrect v4_2| \citep{BR2007}. The Petrosian $r$-band absolute magnitude is 
$^{0.0}M_{r_{Pet}} - 5$ log($h$), which is also $K$-corrected and evolution corrected at $z=0$. We include an extra correction to this absolute 
magnitude of $-0.1$ mag for elliptical galaxies since Petrosian magnitudes underestimate the total flux for these galaxies \citep{Bell+2003,McIntosh+2014}. 
The term $-0.1$ in equation (\ref{logMsBell}) implies a \cite{Kroupa2001} initial mass function (IMF). The systematic error is 0.10--0.15 dex.
The stellar mass estimation used here does not have systematic differences compared to other methods, 
for example, using spectral energy distribution fittings \citep[see][]{Dutton+2011}.

The specific star formation rate (sSFR) is simply defined as the star formation rate divided by the stellar mass. This
quantity has been obtained from the MPA-JHU DR7 catalog,\footnote{Available at http://www.mpa-garching.mpg.de/SDSS/DR7/} 
which corresponds to an updated version of the estimates presented
in \cite{Brinchmann+2004} by using a %spectral
spectrophotometric synthesis fitting model.

The radii used in this work, $R_{deV}$, correspond to the de Vaucouleurs fit scale radius
in the $r$-band (\verb|deVRad_r| parameter in CasJobs). This radius is defined as
the effective (half-light) radius of a de Vaucouleurs brightness profile,
$I(r) = I_0 e^{-7.67 [(r/R_{deV})^{1/4}]}$.

Finally, the luminosity-weighted stellar age is obtained 
from the database of STARLIGHT,\footnote{http://www.starlight.ufsc.br}
through the population synthesis 
method developed by \citet{STARLIGHT+2005} and applied to the SDSS database.

\subsection{Halo masses}
\label{secMh} 

We study how the properties of our isolated elliptical galaxies depend on the host halo mass. %, \mh. 
For this reason, we use the
\citet[][hereafter Y07]{Yang+2007} group catalog, which includes by construction the halo (virial) mass, \mh, of galaxies down 
to some luminosity.
This kind of halo-based group finders have emerged as a powerful method for estimating group halo masses, 
even when there is only one galaxy in the group.  
This method can recover, in a statistical sense, the true halo mass  
from mock catalogs with no significant systematics \citep{Yang+2008}. 

The halo mass in Y07 is based on either the  characteristic stellar mass or the characteristic luminosity in the group. 
We use the halo mass based on the characteristic stellar mass, which
is defined as the sum of the stellar mass of all the galaxies in the halo with $^{0.1}M_r$ - 5 log$(h) \leq$ -19.5, 
where $^{0.1}M_r$ is the $r$-band absolute magnitude with
$K$-correction and evolution correction at $z$ = 0.1. 
They assume a one-to-one relation between the characteristic stellar mass and \mh\ by matching their rank 
orders for a given volume and a given halo mass function.

However, for single-galaxy groups, which are not complete in characteristic stellar mass, the halo mass 
estimates under the assumption of the one-to-one relation mentioned above are already not reliable. 
For this reason, the central galaxies that are fainter than the magnitude limit do not have halo mass estimates, where the 
halo mass lower limit in the Y07 catalog 
%\LEt{FYI:\ You have primarily used US conventions in this paper,  and so I have opted to correct for US style and spelling.} 
is $10^{11.6}$ $\mhalo$.  
We point out that because of the method used in Y07, the estimated halo masses are not measurements of the true 
halo masses, but they are a very good statistical approximation.

\subsection{Redshift and stellar mass limits}
\label{volume}

Since the UNAM-KIAS catalog reaches mainly
out to $z=0.08$, throughout the text we use the redshift range 0 $< z < 0.08$.
Furthermore, in a magnitude limited sample, the minimum detected \ms\ depends on the redshift and on the stellar
mass-to-luminosity ratio, where the latter depends on galaxy colors.
For the SDSS sample and its magnitude limit, \citet[][see also \citealp{Yang+2009}]{vandenBosch+2008} 
%\LEt{Please use a semicolon between the year and the next reference or any term/description. Please check for this throughout.} 
calculated the stellar mass limit at each $z$ above which the sample is complete.
We adopt their  limit as follows: 
\begin{eqnarray} 
%\lefteqn{ \textrm{log}[M_{s,lim}/\mstar] =} \nonumber\\
%&  & {} \frac{4.852 + 2.246 \textrm{ log}(d_L) + 1.123 \textrm{ log}(1+z) %- 1.186z}{1-0.067z}.
\lefteqn{ \textrm{log}(M_{s,lim}/\mstar) } \nonumber\\ 
& & = \frac{4.852 + 2.246 \textrm{ log}(d_L) + 1.123 \textrm{ log}(1+z) - 1.186z}{1-0.067z}.
\label{eq_Mslim}
\end{eqnarray}

Our final sample of isolated ellipticals ($T\le -4$) consists of 89 galaxies using eq. \ref{eq_Mslim} (69 galaxies with halo mass estimates).
For the average redshift  of our sample of isolated E galaxies, $\bar{z}= 0.037$, the corresponding average stellar                   
mass limit is $\log(M_{s,lim}/\mstar)$ = 9.47. 
We notice that only three E galaxies of these 89 have stellar masses smaller than this average mass limit. 

On the other hand, we obtain a final sample of 102 elliptical galaxies in the Coma supercluster according to equation 
(\ref{eq_Mslim}). % with spectroscopic redshifts from CasJobs.
The mass limit completeness at the average redshift of the Coma supercluster galaxies ($\bar{z}=0.023$)
is $\log(M_{s,lim}/\mstar) = 9.0$.

%===========================================
\section{Properties and mass dependences of elliptical galaxies}
\label{secMs}
%===========================================

\subsection{Colors and star formation rates}
\begin{figure}
%\resizebox{\hsize}{!}\includegraphics{Hgi_Ms.eps}
\includegraphics[width=9.8cm]{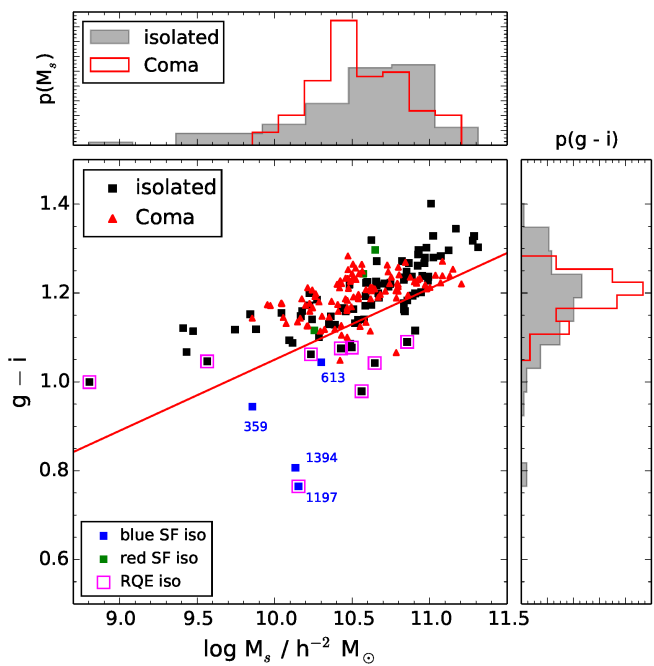}
\caption{
As a function of stellar mass, $g-i$ color  
%where the symbols for the galaxies are the same as Figure \ref{BR_MR}.
for our sample of isolated elliptical ($T \leq -4$) galaxies out to $z=0.08$ (black filled squares). 
In addition, we include a sample of elliptical 
galaxies located in the Coma 
supercluster (red filled triangles). 
The red line shows equation (\ref{eq_gi}) to separate red/blue galaxies.
Blue filled squares and green filled squares show the 
blue and red isolated ellipticals, which are also star-forming galaxies according to 
equation (\ref{eq_sSFR}), respectively. 
The numbers are the ID in the UNAM-KIAS catalog for the former.
Magenta open squares correspond to the recently quenched elliptical (RQE) galaxies following the color-color criterion of Sect. \ref{secRQE}. %\ref{implications}.  
Top and right panels: normalized density distributions of stellar mass and $g - i$ color for isolated elliptical 
galaxies (gray solid histogram) and
elliptical galaxies located in the Coma supercluster (red open histogram), respectively. The integral of each histogram sums to unity.
}
\label{gi_Ms}
\end{figure}

Figure \ref{gi_Ms} shows the $(g-i)$--\ms\ diagram for 
our sample of isolated elliptical %($T \leq -4$) 
galaxies (black filled squares) and for ellipticals in the  
Coma supercluster (red filled triangles).
The red solid line corresponds to the relation found by \cite{Lacerna+2014} to separate red and blue galaxies, specifically 
\begin{eqnarray} 
(g - i)  = 0.16[\textrm{log($M_{s}$)}-10]+1.05  \textrm{,} 
\label{eq_gi}
\end{eqnarray}
%\\
where \ms\ is in units of $\mstar$.
As can be seen from the figure, most of the ellipticals located both in dense and isolated environments are red upon this criterion. 
However, there is a fraction of galaxies with bluer colors. We find that 18 isolated ellipticals ($\approx 20\%$ of the sample) 
are below the red line, some of them far away from this division line. Instead, there are only eight ($\approx 8\%$) ellipticals in Coma that are below the red line, 
and most of them are actually close to it, probably lying in what is called the green valley. Thus, the fraction of blue galaxies is
higher in the isolated environment than in the Coma supercluster. For the blue isolated ellipticals, they become bluer as the mass is lower.  At relatively low masses, $M_s < 10^{10.4}$  $\mstar$, %a 
the blue population %remains that 
only corresponds to the isolated elliptical sample. %and
%they are not seen
%in the Coma cluster.  
At high masses, most of the galaxies are red. 

%%%Fig2 

The top and right panels of Fig. \ref{gi_Ms} show the stellar mass and $g - i$ color distributions, respectively, for isolated elliptical galaxies 
(gray solid histogram) and elliptical galaxies located in the Coma supercluster (red open histogram). %, respectively. 
These normalized density distributions were obtained using the Knuth method for estimating the bin width implemented on \verb|astroML|\footnote{http://www.astroml.org/} \citep{astroMLText}, which is also able to recognize substructure in data sets.
The $g-i$ mean and median values of isolated and Coma supercluster ellipticals are similar, see Table \ref{tabla_full}, although
the distribution is slightly narrower in the %latter case
case of Coma galaxies
%\LEt{This is possibly ambiguous. If you mean for "Coma supercluster ellipticals", then it's fine. If not, please specify.} 
(see also the difference among the 16th and 84th percentiles). 
Some isolated elliptical galaxies seem to be more massive than the most massive elliptical galaxies in the dense environment  
(e.g., NGC 4889 and NGC 4874 of Coma cluster), though the differences are within the mass uncertainties of 0.15 dex. 
%We use spectral synthetic models to check that our mass estimates for these galaxies are consistent with other stellar mass estimations, 
%such as those from the MPA-JHU DR7 catalog.
We checked that our mass estimates for these galaxies are consistent with mass estimations based on spectral synthetic models from the MPA-JHU DR7 catalog.
%\LEm{The suggested change seemed to contradict our intended meaning. We rephrase the sentence to make it clearer.}

%%% TABLA full
\begin{table*}
\caption{Physical properties of isolated ellipticals and E galaxies in the Coma supercluster. }
  \centering
\begin{tabular}{c c c c c c c c c c c c c}
\hline\hline
         %& &      & UNAM-KIAS&        &      &      & &      &          & Coma   &      & \\
& \multicolumn{6}{c}{isolated}  & \multicolumn{6}{c}{Coma}  \\
\cline{3-7}
\cline{9-13}
property & & mean & $\sigma$ & median & 16th & 84th & & mean & $\sigma$ & median & 16th & 84th\\
%(1)    &  (2)   & (3) & (4) & (5) & (6) \\
\hline
log(\ms)\tablefootmark{a} & &10.55 & 0.46     & 10.62  & 10.17& 10.95& &10.55 & 0.29     &   10.51  & 10.24& 10.84\\
$g - i$  & &1.18  & 0.10     & 1.19   & 1.09 & 1.28 & & 1.20 & 0.05     & 
1.20   & 1.14 & 1.24 \\ 
log(sSFR)\tablefootmark{b}& &-11.88& 0.52     & -12.00 &-12.27&-11.61& &-12.11& 0.30     &
-12.17 &-12.36& -11.87 \\
$^{0.0}M_R$& & -21.87 & 1.06 & -22.02 &-22.77&-20.95& &-21.78& 0.69     &
-21.71 &-22.51& -21.05 \\
$B-R$      & & 1.62   & 0.10 & 1.63   & 1.53 & 1.72 & & 1.64 & 0.05     &
1.64   & 1.58 & 1.68 \\
log(R$_{deV}$)\tablefootmark{c} & &0.53& 0.25 & 0.55   & 0.29 & 0.76 & & 0.50 & 0.20     &
0.50   & 0.33 & 0.69 \\
log(age)\tablefootmark{d}       & &9.63& 0.30 & 9.67   & 9.42 & 9.86 & & 9.72 & 0.20     &
9.74   & 9.59 & 9.88 \\
\hline
\end{tabular}

\tablefoot{For each property, the columns correspond to the mean, its standard deviation, the median, the 16th, and 84th percentiles.\\% of each property.\\
\tablefoottext{a}{log$_{10}$ of the stellar mass in units of $\mstar$.}
\tablefoottext{b}{log$_{10}$ of the specific star formation rate in units of yr$^{-1}$.}
\tablefoottext{c}{log$_{10}$ of the radius of a de Vaucouleurs fit in the $r$-band in units of kpc ($h=0.7$).}
\tablefoottext{d}{log$_{10}$ of the light-weighted stellar age in units of yr.} 
}
\label{tabla_full}
\end{table*}

\begin{figure}
\includegraphics[width=9.8cm]{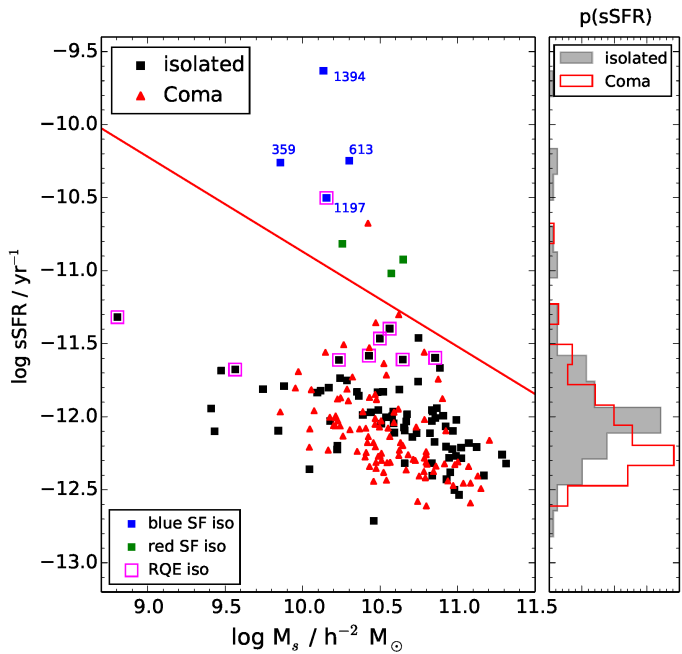}
\caption{Specific star formation rate (sSFR) as a function of the stellar mass.
The symbols for the galaxies are the same as  Fig. \ref{gi_Ms}.
The red line shows equation (\ref{eq_sSFR}) to separate star-forming and passive galaxies (above and below the line, respectively). 
Right panel: normalized density distribution of sSFR for isolated elliptical galaxies (gray solid histogram) and
elliptical galaxies located in the Coma supercluster (red open histogram). The integral of each histogram  sums to unity.
}
\label{sSFR_Ms}
\end{figure}

Figure \ref{sSFR_Ms} shows sSFR as a function of stellar mass. 
We include as a red solid line the relation found by \cite{Lacerna+2014} to separate passive and SF
%star-forming (hereafter SF) 
galaxies, i.e.,%
\begin{eqnarray} 
\textrm{log(sSFR)} = -0.65[\textrm{log($M_{s}$)}-10]-10.87  \textrm{ ,} 
\label{eq_sSFR}
\end{eqnarray}
%\\
where \ms\ is in units of $\mstar$ and sSFR is in units of yr$^{-1}$.
According to this relation, galaxies located above and below this line are considered  
SF and passive galaxies, respectively.  Most of our isolated ellipticals 
are passive galaxies in terms of their star formation.
In Figs. \ref{gi_Ms}, \ref{sSFR_Ms}, and the  figures that follow, we plot with a blue (green) solid square those ellipticals that are blue (red) 
and SF galaxies; the magenta open squares highlight the `recently quenched ellipticals' to be described below. 
There are four blue SF ellipticals and three red SF galaxies. In total, we have seven SF isolated ellipticals 
($\approx 8\%$ of all the isolated ellipticals). In Sect. \ref{secBPT}, 
we see that at least one blue SF galaxy could be classified as AGN/LINER 
in the BPT diagram \citep[][]{BPT1981}
and in Appendix \ref{Ap_BlueSF} that other blue SF galaxy could be an AGN because of its broad component in H${\alpha}$.
% rather than SF galaxy, leaving a fraction of only $\lessapprox 7\%$. 
On the other hand, nearly all the E galaxies from the Coma supercluster (red filled triangles) are passive objects according to the above criterion. 
There is only one ($\la 1\%$) elliptical in Coma that would be a SF galaxy. 

%%%Fig3 

In general, the distribution of isolated and Coma ellipticals in the sSFR--$M_s$ plane seems to be not too 
different, except for the small fraction of isolated ellipticals at masses $10^{9.7}< M_s < 10^{10.7}$ $\mstar$ 
with relevant signs of star formation activity.
The right panel of Fig. \ref{sSFR_Ms} shows the sSFR distributions for  isolated elliptical galaxies (gray solid histogram) 
and elliptical galaxies in the Coma supercluster (red open histogram). The latter are slightly more passive, by $\sim0.2$ dex according to 
the mean and median values reported in Table 1.

\subsection{The color--magnitude diagram}

\begin{figure}
\includegraphics[width=9.8cm]{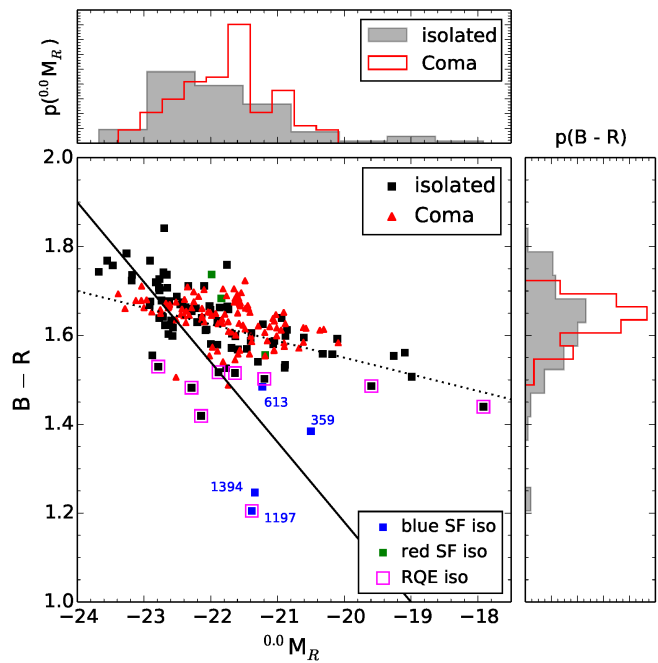}
\caption{$B-R$ color as a function of the absolute magnitude in the $R$-band ($h=0.7$, $K$-corrected at $z = 0$). The symbols for the galaxies are the same as Fig. \ref{gi_Ms}.
The linear fit of \cite{Niemi+2010} for their (nonisolated) elliptical galaxies from a semianalytic model is shown as a dotted line. 
The fit for their simulated isolated elliptical  galaxies is shown as a solid line.
Top and right panels: normalized density distributions of absolute magnitude and $B-R$ color for isolated elliptical 
galaxies (gray solid histogram) and elliptical galaxies located in the Coma supercluster (red open histogram), respectively. The integral of each 
histogram sums to unity.
}
\label{BR_MR}
\end{figure}

It is well known that, in general, more luminous galaxies tend to have redder colors.
In Fig. \ref{BR_MR}, as an example of the color--magnitude diagram (CMD), we plot the $B - R$ color against $R$-band 
absolute magnitude for our samples of isolated and Coma E galaxies. The symbols for the E galaxies are the same 
as in Fig. \ref{gi_Ms}. The trend that galaxies are redder as their luminosity increases  %\LEt{Or, perhaps, "galaxies are redder as their luminosity increases".} 
is followed by 
both isolated and Coma E galaxies.
This means that ellipticals exhibit roughly a similar CMD regardless of the environment. However, as in the case of the 
color--\ms\ diagram, there is a small fraction of isolated ellipticals that follow a different trend, toward bluer colors as magnitudes 
are fainter.  They show a steeper color--magnitude relation than the rest of the ellipticals, giving rise to a branch that systematically 
detaches from the red sequence. The blue SF galaxies identified in the previous figures (blue filled squares) are namely at the end of this branch. 

%%%Fig4

The right panel of Fig. \ref{BR_MR} shows the $B - R$ color distribution for isolated E galaxies (gray solid) and Es in the 
Coma supercluster (red open). The mean and median $B - R$ colors (reported in Table \ref{tabla_full}) are 
only slightly different between both samples. Elliptical galaxies in Coma are on average redder by $\sim 0.02$ mag than those in isolation,
but the distribution of the former is narrower than that of the latter (see also the 16th and 84th percentiles). 
The top panel of this figure corresponds to the distribution of the absolute magnitude in the $R$-band. The mean and median 
values are reported in Table \ref{tabla_full}. A slight bias is seen in the distribution toward brighter isolated ellipticals than in Coma.

\subsection{Recently quenched ellipticals and stellar ages}     
\label{secRQE}

\citet{McIntosh+2014} introduced some criteria to define the population of ellipticals that have been quenched recently.
These recently quenched ellipticals (hereafter RQE) are mostly blue, have light-weighted stellar ages shorter than 
3 Gyr \citep[according to age determinations by][]{Gallazzi+2005}, and lack of detectable emission for star formation. 
Such galaxies clearly experienced a recent quenching of star formation
and are now transitioning to the red sequence.  \citet{McIntosh+2014} argue that these RQEs ``have recent star formation histories 
that are distinct from 
similarly young and blue early-type galaxies with ongoing star formation (i.e., rejuvenated early-type galaxies)'', and given their 
supra solar metallicities,
they ``are consistent with chemical enrichment from a significant merger-triggered star formation event prior to the quenching''. 
The authors conclude
that RQEs are strong candidates for `first generation' ellipticals formed in a relatively recent major spiral-spiral merger. 
\citet{McIntosh+2014} found 
an empirical criterion in the $(u-r)-(r-z)$ diagram to select RQEs.

\begin{figure}
\includegraphics[width=8.8cm]{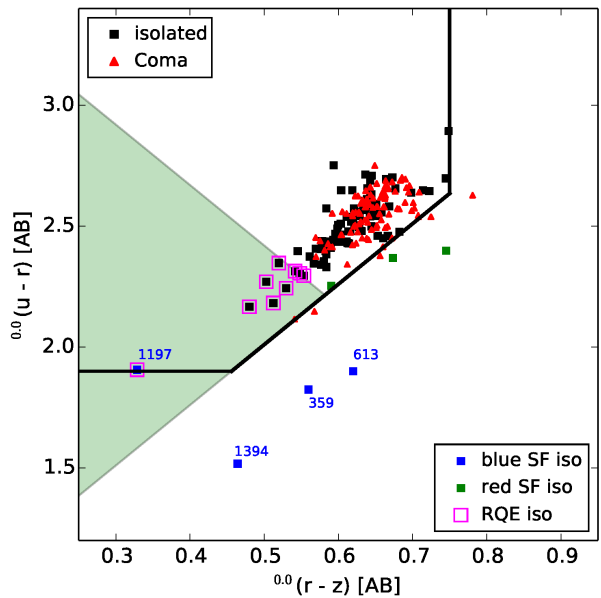}
\caption{Color--color diagram to identify recently quenched elliptical (RQE) galaxies. The colors are in AB magnitudes and $K$-corrected 
at $z=0$.
The symbols for the galaxies are the same as Fig. \ref{gi_Ms}.
The thick black lines enclose the non-star-forming region of \citet{McIntosh+2014} based on \citet{Holden+2012}. The RQE selection region 
is empirically defined as the green shaded triangle by \citet{McIntosh+2014}. Nine isolated elliptical galaxies are  inside this region  and
 are shown in magenta open squares.
}
\label{RQE}
\end{figure}

Figure \ref{RQE} shows the $(u-r)-(r-z)$ diagram for the isolated and Coma E galaxies. The green shaded triangle is the empirical 
region where RQEs lie according to \citet{McIntosh+2014}. There are nine ($\approx 10\%$) isolated E galaxies that are RQEs according to this criterion; 
they are shown with magenta open squares in Fig. 5 and in other figures. Among these nine ellipticals, seven are blue and two are red, according 
to equation (\ref{eq_gi}), as can be seen in Fig. \ref{gi_Ms}, and eight are passive, according to equation (\ref{eq_sSFR}), as can be 
seen in Fig. \ref{sSFR_Ms}.  In Sect. \ref{secBPT} we see that the (marginal) RQE, which is a blue SF galaxy 
(highlighted with a blue solid  square; UNAM-KIAS 1197), can instead be  classified as a LINER according to the BPT diagram.
There are %not 
no isolated RQEs more massive than 
$\ms\approx 7\times 10^{10}$ $\mstar$ (see, e.g., Fig. \ref{gi_Ms}), and they reside in haloes of masses 
$\lesssim 6\times 10^{12}$ $\mhalo$ (see Fig. \ref{BR_Mh} below), confirming that the most massive ellipticals were quenched long ago. 
The two RQEs classified as red galaxies are in fact the least massive of the nine RQEs; see Fig. \ref{gi_Ms}. 

%%%Fig5

In the case of the ellipticals in the Coma supercluster, there is only one RQE candidate although it lies in the lower border limit. Thus, while in the
Coma supercluster there are virtually no RQEs (even those that are blue in Fig. \ref{gi_Ms} seem to have been quenched relatively early), 
RQEs are $\approx 10\%$ among isolated E galaxies.  
This environmental difference is consistent with \citet{McIntosh+2014}, where the authors find that the vast majority of their RQEs 
are central in smaller groups ($\mh\lesssim 3\times 10^{12}$ $\mhalo$), i.e., they do not reside in high-density environments. 

The black solid lines in Fig. \ref{RQE} delimit the region of spectroscopically quiescent, red-sequence early-type galaxies 
according to \citet{Holden+2012} and extended  to $(u-r)=1.9$ in \citet{McIntosh+2014}. Galaxies outside this boundary 
are defined as pure star-forming (BPT \ion{H}{ii} emission). 
Most of the ellipticals in Coma (triangles) and isolated ellipticals (squares) qualify as non-SF systems in 
this diagram. 
There is a rough agreement between those galaxies defined as SF in Figs. \ref{sSFR_Ms} and \ref{RQE}.
Blue SF isolated E galaxies (blue squares with numbers corresponding to their ID in the UNAM-KIAS catalog)
are located far away from the non-SF region, whereas red SF isolated E galaxies (green squares) are close to the border. 

\begin{figure}
\includegraphics[width=9.8cm]{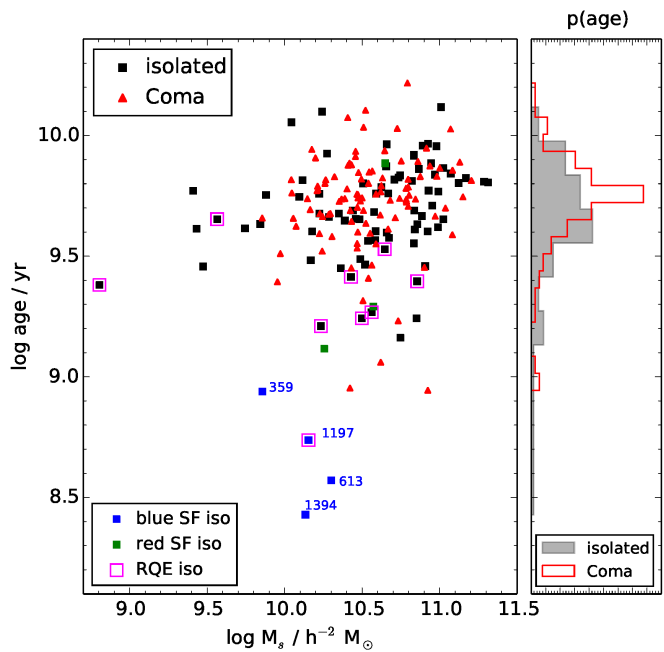}
\caption{Luminosity-weighted stellar age %(\textbf{$h$ = 0.7?}) 
as a function of stellar mass. 
The symbols for the galaxies are the same as Fig. \ref{gi_Ms}.
Right panel: normalized density distribution of the luminosity-weighted stellar age for isolated elliptical galaxies (gray solid histogram) 
and
elliptical galaxies in the Coma supercluster (red open histogram). The integral of each histogram sums to unity.
}
\label{ageF_Ms}
\end{figure}

In contrast to properties such as color and size (see Sect. \ref{sec_sizes}), the luminosity-weighted stellar age does not show 
a significant dependence on 
stellar mass as shown in Fig. \ref{ageF_Ms}. From this figure it is clear that the blue SF isolated ellipticals are the youngest 
galaxies with luminosity-weighted stellar ages $\lesssim 1$ Gyr. This could indicate recent processes of star formation for this particular 
class 
of objects. The right panel shows the stellar age distribution of isolated ellipticals (gray solid histogram) and elliptical galaxies 
located in the Coma supercluster (red open histogram). The mean and median values are reported in Table \ref{tabla_full}. Ellipticals
in Coma appear to be older than isolated ellipticals in general ($\sim 1$ Gyr older), although this difference is within 
the statistical uncertainties.   
%HECTOR: IVAN un test estadistico para ver diferencias/semejanzas?
%IVAN: creo que con los valores reportados en Tabla 1 es suficiente.

%%%Fig6 

\subsection{Galaxy sizes}
\label{sec_sizes}

\begin{figure}
\includegraphics[width=9.8cm]{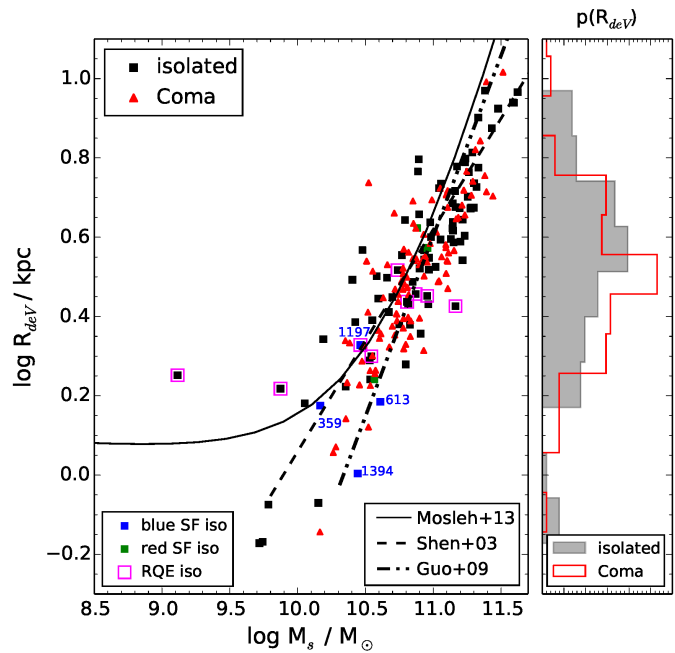}
%\resizebox{\hsize}{!}{\includegraphics{H_RdeV_Ms.eps}}
\caption{
Effective radius of a de Vaucouleurs fit, R$_{deV}$, %containing 50\% of Petrosian flux 
in the $r$-band as a function of stellar mass ($h = 0.7$). 
The symbols for the galaxies are the same as Fig. \ref{gi_Ms}.
We also include the mass--size relations of early-type galaxies by \citet[][dashed line]{Shen+2003}, \citet[][dot-dot-dashed line]{Guo+2009} and 
\citet[][solid line]{Mosleh+2013}.
%\LEt{Please correct formatting to place a semicolon between the year and subsequent term throughout.}
Right panel: normalized density distribution of the radius for isolated elliptical galaxies (gray solid histogram) and
elliptical galaxies in the Coma supercluster (red open histogram). The integral of each histogram sums to unity.
}
\label{R50_Ms}
\end{figure}

Figure \ref{R50_Ms} shows the effective
radius in the $r$-band for the different samples of elliptical galaxies as a function of stellar mass. The general trend that massive
galaxies are bigger with no apparent dependence on environment is observed. Blue SF isolated ellipticals are smaller than 2.5 kpc. %indicating they are compact in nature. 
% IVAN: otra vez, no quiero confundir con literatura reciente sobre isolated compact E galaxies (e.g. Chilingarian & Zolotukhin 2015) donde el radio efectivo es menor a 0.6 kpc.
In the case of RQEs, they have R$_{deV} \lesssim 3.5$ kpc. 
The right panel of Fig. \ref{R50_Ms} shows the radius distribution of the isolated and Coma supercluster ellipticals 
(gray solid and red open histograms, respectively). We see that the radius distribution of 
Coma and isolated ellipticals are similar. If any, the former is slightly shifted to smaller radii with respect to the latter.  
The mean and median values of the radii are reported in Table \ref{tabla_full}. 
On average the radius for both samples are similar within the uncertainties,
although the distribution of E galaxies in Coma is narrower than that of isolated ellipticals (e.g., see the 16th and 84th percentiles).

%%%Fig7 

In Fig. \ref{R50_Ms} we also include the size--mass relation of early-type galaxies ($n > 2.5$) reported in 
\citet[][]{Shen+2003, Shen+2007} as a dashed line. They use the half-light radius in the $z$-band to define the size of
 their galaxies with $z \geq 0.005$.  
The dot-dot-dashed line shows the size--mass relation found by \citet{Guo+2009} for early-type galaxies 
(usually $n > 3.5$) at $ z \leq 0.08$. %,  $ which was 
They define the size as the half-light radius in the $r$-band.
The solid line is the relation found by \citet{Mosleh+2013} for their sample of early-type galaxies between $0.01 < z < 0.02$. 
The morphology of these galaxies was obtained from the Galaxy Zoo Catalogue \citep{GZC+2011}; the size is also defined  as 
the half-light radius in the $r$-band. Despite the fact that different size,  morphological definitions, and samples were 
used in each one of these works, we find a qualitative good agreement among their studies and this work at masses $\ms > 3 \times 10^{10}$ $\msun$. 
The best agreement is obtained with the result by \citet{Guo+2009}. Their sample is more similar to ours 
(redshift limit, size definition, and more stringent criteria to select early-type systems). At lower masses, \citet{Mosleh+2013} 
find a flatter size--mass relation. This is roughly followed by some of our low-mass isolated galaxies, although other 
low-mass ellipticals follow an extension of the trend exhibited by the massive %isolated 
galaxies. It seems that the
size--mass relation is different or not unique at the low-mass end.

\subsection{BPT diagram}
\label{secBPT}

A nuclear classification was carried out from diagrams in the optical diagnosis initially introduced by \citet[][BPT]{BPT1981}
and redefined by \citet{VeilleuxOsterbrock1987}. The BPT diagrams
%\LEt{Please avoid beginning sentences with abbreviations, acronyms, numbers in figures, and the like. Please check for this throughout the paper.} 
have been massively
used  for discriminating between different mechanisms of ionization and production of emission lines, separating 
the production mechanisms either by photoionization by massive OB stars (typical of regions with star formation) 
or nonstellar sources such as the presence of AGN in the core. 
We have cross-matched our sample of isolated and Coma supercluster E galaxies with the database of STARLIGHT, %\textbf{\footnote{http://www.starlight.ufsc.br}} 
where the stellar population synthesis analysis developed
by \citet{STARLIGHT+2005} was applied to SDSS galaxies.

\begin{figure}
\includegraphics[width=8.8cm]{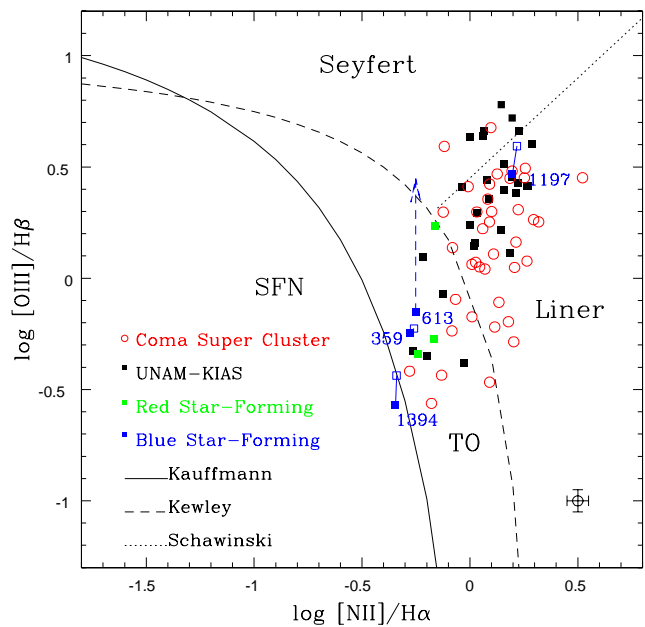}
\caption{
BPT diagram for the isolated elliptical galaxies (black filled squares) of the UNAM-KIAS catalog and elliptical galaxies in the Coma supercluster (red open circles) with estimates from the STARLIGHT database. The typical uncertainty is indicated with the black open symbol with error bars.
Blue and green solid points represent the blue SF and red SF isolated E galaxies, respectively. 
Open blue squares indicate particular measurements for blue SF isolated galaxies using the method described in Appendix \ref{Ap_Spec}.
Solid blue lines just connect the two different measurements for each blue SF elliptical, where the numbers correspond to their ID in the UNAM-KIAS catalog.
Galaxies are classified as star-forming nuclei (SFN), transition objects (TO), or AGN (Seyfert/LINER) according to the relations of \citet[][solid line]{Kauffmann+2003},  
\citet[][dashed line]{Kewley+2001}, and \citet[][dotted line]{Schawinski+2007}. 
The dashed arrow indicates that UNAM-KIAS 613 %can be classified as
is rather a Seyfert 1.8 because of its broad components in H$\alpha$ and H$\beta$ (see details in Appendix \ref{Ap_BlueSF}). 
}
\label{BPT}
\end{figure}

Figure \ref{BPT} shows the BPT diagram for the isolated E galaxies with reported lines in the STARLIGHT database (solid squares). 
Those with reported lines %represent
correspond to 35 out of 89 ($\approx 39\%$) isolated E galaxies. 
%\LEt{If this is not correct, please fix to be more specific.}
The rest are probably quenched and/or without an AGN at present day. 
Out of the 35 galaxies, 24 are AGNs (19 LINERS and five Seyferts) and ten are transition 
objects (TOs).\footnote{The interpretation of TO objects as SF-AGN composites is not clear. For example, \citet{McIntosh+2014} 
discuss that these galaxies may be neither SF nor an AGN, rather their emission may be dominated by the same non-nuclear 
ionization sources as many LINERs.}
Only one galaxy (UNAM-KIAS 1394) is a SF nuclear (SFN) object. This galaxy is also a blue SF object as defined by us. 
Other two blue SF isolated Es are TOs, and one blue SF galaxy seems to be actually a LINER (UNAM-KIAS 1197). 
In Appendix \ref{Ap_BlueSF} and \ref{Ap_Spec}, we describe a more detailed
spectroscopical analysis of these four blue SF isolated ellipticals, using lines with  S/N$>$7. Our results (blue open squares) are 
similar to those from the STARLIGHT database (blue solid squares). %plotted in Fig.  \ref{BPT}, 
%excepting for UNAM-KIAS 613 that lies in the border of the Seyfert region. 
In the case of UNAM-KIAS 613, we find that this galaxy is actually a Seyfert 1.8  because of its broad components in H${\alpha}$ and H${\beta}$ (see details in Appendix \ref{Ap_BlueSF}). Therefore, 25 isolated Es correspond to AGNs, nine are TOs and one galaxy is a SFN.

%%% Fig8

In the case of the E galaxies located in the Coma supercluster (red open circles), 39 out of 102 ($\approx 38\%$) have reported lines in the STARLIGHT database. The rest are probably quenched and without an AGN by today. Twenty-nine galaxies are LINERs, two are Seyferts and eight are TOs. 

In general, there is a similar fraction of isolated and Coma supercluster ellipticals that present ionization emission lines
associated with star formation or AGN activity. Among these, the fraction of those classified as SFN/TO is 
slightly larger for the isolated ellipticals than for the Coma ellipticals (11\% vs 8\%).  Thus, environment seems to have
a weak influence on the fractions of ellipticals with a mixture of emission coming from a
circumnuclear SF region and central low-luminosity AGN in the BPT diagram.

%============================= 
\section{Interpretation and implications}    
\label{implications}
%============================= 

The results presented above are consistent with several previous works on early-type galaxies mentioned in the Introduction.  
Our sample is different from previous samples in that it selects very isolated environments and refers only to morphologically 
well-defined elliptical galaxies ($T\leq -4$) in a wide mass range. This allows us to probe the morphological transformation 
and quenching of the star formation as a function of mass, hopefully free of environmental effects.

The work by \citet{Vulcani+2015} is the closest  to the present paper.  Using an automatic tool designed to attempt to reproduce
the visual classifications \citep[MORPHOT;][]{Fasano+2012}, these authors made an effort to distinguish pure E galaxies from S0/Sa 
galaxies in a sample complete above $\log$($\ms/\msun$)$=10.25$ at their redshift upper limit $z=0.11$. For those 
galaxies that \citet{Vulcani+2015} call singles (no neighbors with a projected mutual distance of 0.5 $h^{-1}$ Mpc and a redshift 
within 1500 km s$^{-1}$), $24\pm3\%$ are ellipticals. Out of them, $\approx 83\%$ are red according to their definition. The rest 
are blue and green (roughly 13\% and 4\%, respectively).  In our sample, the fraction of blue isolated ellipticals (we separate
them only  into  blue and red galaxies) above $\log$($\ms/\mstar$)$=9.94$ (10.25 for $h=0.7$) is 
$\approx 21\%$, which is slightly larger than in \citet{Vulcani+2015}, but yet within their error bars.

%%differences between isolated and Coma E's.
According to our analysis, the colors, sSFR, sizes, and luminosity-weighted ages of very isolated E galaxies are only slightly 
different from those in the Coma supercluster (see Table \ref{tabla_full} for mean and median values). %and KS tests????). 
In general, in both environments the fractions of blue and SF ellipticals are low. However, these fractions
are larger for the isolated ellipticals (approximately 20\% and 8\%, respectively) than for the Coma supercluster 
(approximately 8\% and $\la 1\%$, respectively), as seen in Figs. \ref{gi_Ms} and \ref{sSFR_Ms}. Moreover, these fractions
deviate in a different way from the main red/passive sequences as a function of mass. The main difference is that the
blue/SF isolated ellipticals deviate more from the main trends %as smaller they are 
toward smaller masses
down to
$\log$($\ms/\mstar$) $\approx 9.85$ (at lower masses, there are no blue nor SF E galaxies). %\LEt{"as smaller they are.." is ambiguously worded. Perhaps, "because at their smallest they are as low as"?} 
In the case of the small fraction of blue %SF 
ellipticals in the Coma supercluster, they deviate from the main trend only 
moderately, and below $\log$($\ms/\mstar$) $\approx 10.4$ there are no blue %SF 
ellipticals. 
Thus, in a high-density environment, only a small fraction of intermediate-mass galaxies are slightly away from the red sequence today, so that they seem to have intermediate-age stellar populations. 
The situation is not too different for the very isolated ellipticals,
although as mentioned above, the fraction of blue objects is larger and, especially  at intermediate masses, some of them deviate 
significantly from the red sequence and are actually SF galaxies. An interesting question is why these very isolated galaxies
remain blue and star forming after their morphological transformation: Is that because this transformation happened recently from
gaseous mergers or because they accreted gas, suffering a rejuvenation process?  We discuss 
this question in Sect. \ref{blueSF_Es}.

%association of the morphological transformation mechanisms with the quenching of SF
A main result to be discussed now is that E galaxies in the local Universe are mostly red and dead (passive), that is, even those that are very isolated.
It seems that the mechanisms responsible for the morphological transformation of galaxies produce efficient quenching of star formation
and depletion of the cold gas reservoir, both for isolated and cluster E galaxies, and environment is not the 
most effective (or only) mechanism of quenching, although it is expected to play some role (see below Sect. \ref{quenching}).

\begin{figure*}
\includegraphics[width=8.8cm]{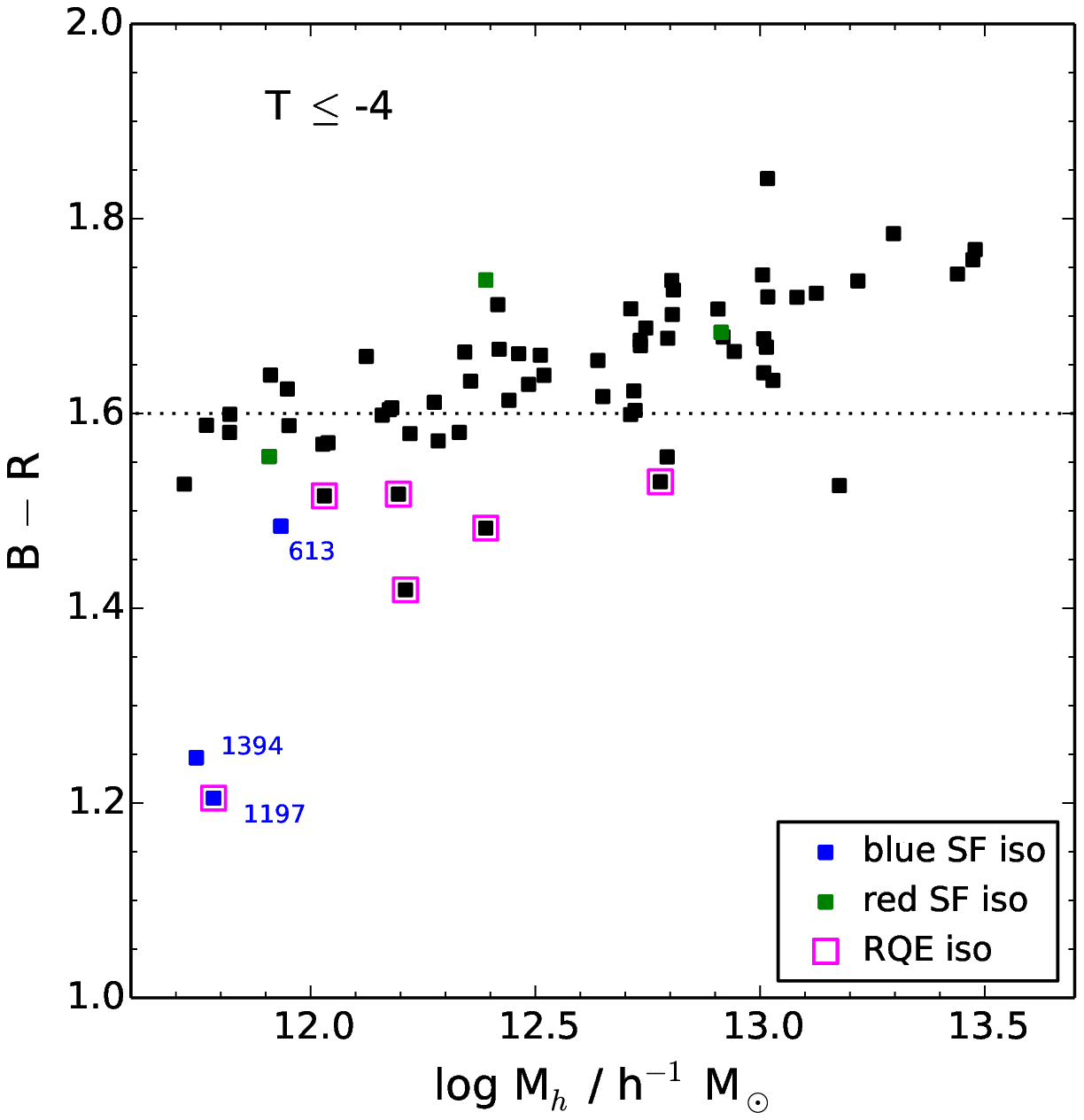}
\includegraphics[width=8.8cm]{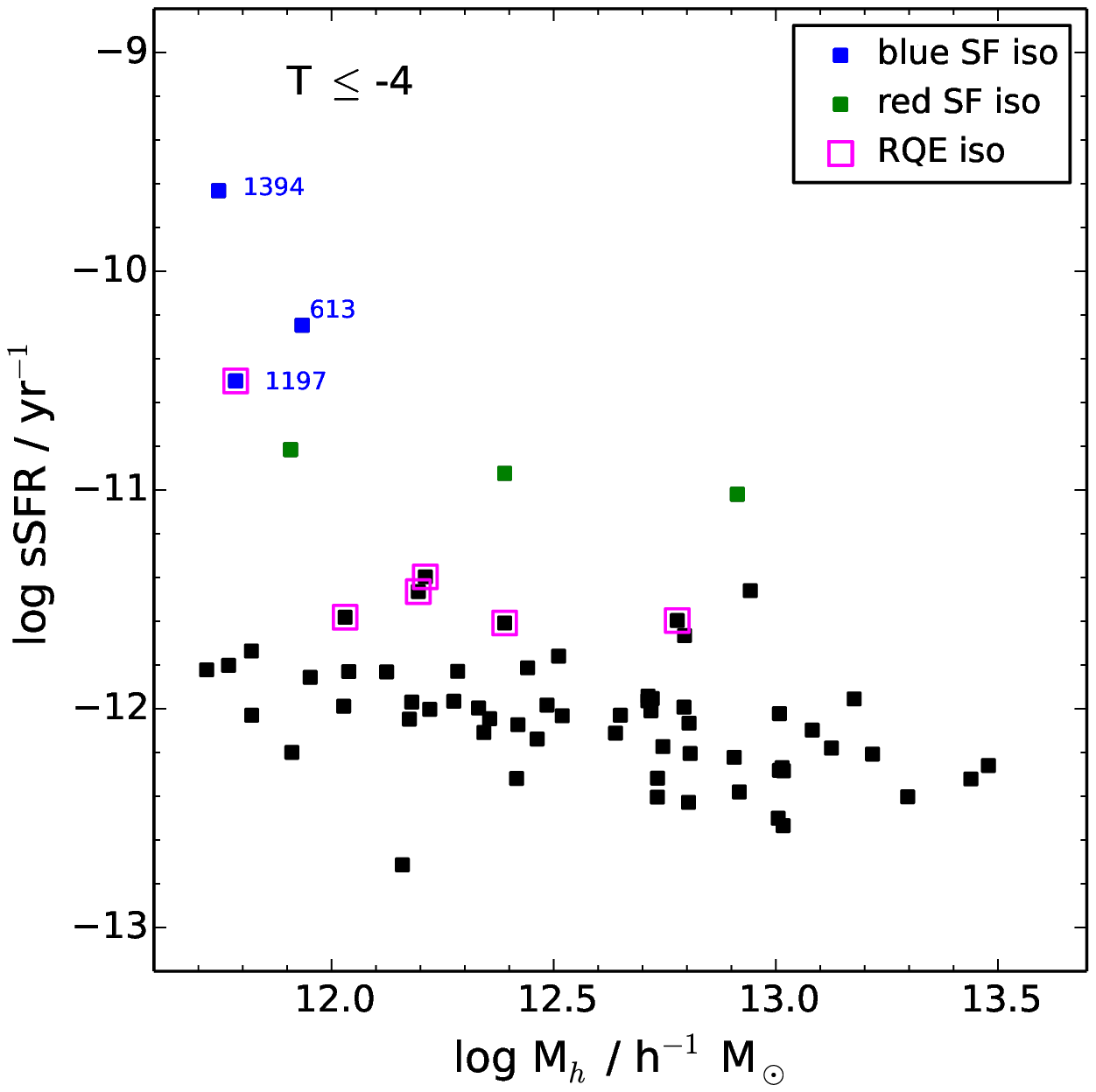}
\caption{
$B-R$ color (left panel) and sSFR (right panel)
as a function of the halo mass of isolated elliptical ($T \leq -4$) galaxies. 
Blue and green squares show those blue %faint 
and red star-forming isolated galaxies of the previous figures with halo mass estimates, respectively.
The numbers are the ID in the UNAM-KIAS catalog for the blue SF galaxies.
Magenta open squares correspond to the isolated RQEs. 
The dotted line in the left panel is a visual approximation for the trend of simulated elliptical galaxies in \cite{Niemi+2010}.
}
\label{BR_Mh}
\end{figure*}

In more detail, as can be seen in Figs. \ref{gi_Ms} and \ref{BR_MR}, both isolated and supercluster ellipticals follow the overall trend that they are on average redder as they %become
are more massive (luminous). Moreover, above $\ms\approx 8\times 10^{10}$ $\mstar$ there are no blue or SF ellipticals at all in both the supercluster and isolated samples, and the luminosity-weighted ages 
are older than 4 Gyr for almost all of them. This confirms that massive E galaxies assembled their stellar populations early, remaining 
quenched since those epochs \citep[cf.][]{Thomas+2005,Schawinski+2009,Kuntschner+2010,Thomas+2010}.  As we go to lower masses, notwithstanding 
the environment, ellipticals tend to have on average bluer colors than the more massive E galaxies.
This mass downsizing behavior for E galaxies is also seen in the case of halo mass.
The left panel of Fig. \ref{BR_Mh} shows the $B-R$ color as a function of \mh\ for the isolated E galaxies %, which t have
with an estimate of 
their halo (group) 
mass according to Y07. %\citet[][see \S\S \ref{secMh}]{Yang+2009}. 
With a large scatter, we find that $B-R\propto 0.12\log\mh$. This trend, 
within the context of the merger-driven (morphological) quenching mechanism, might be explained by more gaseous and/or 
later mergers as the system %becomes
is less massive; actually, observations show that lower mass galaxies %are on average more
%gaseous \citep[see e.g. for a recent compilation][and more references therein]{Calette+2015}. 
%IVAN: Existe referencia?
have on average higher gas fractions \citep[e.g.,][]{Avila-Reese+2008,Papastergis+2012,Calette+2015,Lehnert+2015}.
However, even if the merger is late and gaseous, \emph{the quenching seems to be so efficient and rapid 
that most of the low-mass ellipticals, both isolated and in the Coma supercluster, already transited  to the red 
sequence by $z\sim 0$}.
%\LEt{According to journal style, please avoid italics used for emphasis. Please check for this throughout and correct as italics is used frequently.} 
%\LEm{Can we use "\textbackslash emph" or other alternative for emphasis in the text?}
Even more, their present-day sSFR's are very low, close to those of the massive 
galaxies or haloes (see Fig. \ref{sSFR_Ms} and right panel of Fig. \ref{BR_Mh}). 

%%%Fig9

%the isolated and Coma Es in more detail regarding their quenching properties
Is it possible to evaluate in a more quantitative way whether the quenching of those E galaxies that are passive today happened 
early or recently? The very red colors of most massive ellipticals ($\ms\gtrsim 10^{11}$ $\mstar$) suggest that these galaxies 
did not have active star formation since at least 1-2 Gyr ago. However, those (less massive) passive E galaxies with bluer colors and 
sSFRs slightly higher than the average could have been quenched more recently.  As described in Sect. \ref{secRQE}, in the 
plot presented in Fig. \ref{RQE}, we can select those RQE galaxies. At face value, our results show that the ellipticals in an 
environment such as the Coma supercluster suffered the quenching of star formation, and likely the previous morphological transformation, early, so that at $z\sim 0$ almost all of them are passive and red; the small fraction that lies below the red sequence 
($\approx 8\%$ in Fig. \ref{gi_Ms}, likely in the green valley) are not RQEs excepting one marginal case, i.e. these ellipticals
started to be quenched long ago but they are just in the process of transition to the red sequence. 

In the case of isolated ellipticals, among those that are blue but passive galaxies ($\sim$15\%), half are RQEs 
(see Figs. \ref{gi_Ms} and \ref{RQE}). The fact that in the isolated environment, among the blue passive ellipticals there 
is a significantly larger fraction of RQEs than in the Coma supercluster, suggests that the transition to the red sequence should
be faster for isolated ellipticals than for supercluster ones. For the blue passive isolated E galaxies, we can even 
see a trend. The RQEs are bluer than those that were quenched not recently; most of the latter already lie  very close to the 
red sequence (see Fig. \ref{gi_Ms}). Instead, almost all of the (few) blue Coma ellipticals were quenched not recently but 
they did not yet arrive to the red sequence. Moreover, for masses below log($\ms/\mstar$) $\sim 10.5$,
%\textbf{$\log$($\ms/\mstar$)$\sim 10$, 
the isolated RQEs are close or already in the red sequence (excepting the peculiar case of the bluest galaxy to be discussed below;
see Fig. \ref{gi_Ms}).
Therefore, they quenched and reddened very fast. 
In general, we see that \emph{as less massive the isolated RQEs  are, the faster they seem to have been reddened}. 
What produce the cessation of star formation and gas accretion so efficiently in isolated low-mass E galaxies?

\subsection{Quenching mechanisms of E galaxies}
\label{quenching}

Our results are consistent with the conclusions by \citet{Schawinski+2014} for early-type galaxies
in general. These authors propose that a major merger could simultaneously transform the galaxy 
morphology from disk to spheroid and cause rapid depletion of the cold gas reservoir with a 
consequent quenching of star formation (morphological quenching). As a result of the drop in star formation, 
the galaxy moves out of the blue cloud, into the green valley, and to the red sequence as fast
as stellar evolution allows. The authors estimate that the transition process in terms of galaxy
color takes about 1 Gyr for early-type galaxies; this time is much longer for late-type galaxies,
i.e., they undergo a much more gradual decline in star formation.  The rapidity of the gas reservoir 
destruction in E galaxies should be due to a very efficient star formation process
and strong %SN- 
supernova-
and/or AGN-driven feedback (winds, ionization, etc.). 

According to our results, on the one hand, isolated and Coma supercluster E galaxies in general share the same loci in the
color--\ms, color-magnitude, and sSFR--\ms\ diagrams, and show evidence of rapid transition to the 
red sequence after they were quenched, especially the low-mass isolated ellipticals.
On the other hand, the radius--mass relation of supercluster and isolated E galaxies are roughly similar and
follow roughly the relation determined for early-type galaxies from large samples (Fig. \ref{R50_Ms}). 
Moreover, as is well known, ellipticals in general are more concentrated than late-type galaxies. 
%This is evidenced by strong dissipative processes at the basis of the origin in most of these galaxies, 
This makes evident the presence of strong dissipative processes at the basis of the origin in most of these galaxies, 
which are happening in the same way both for the isolated and cluster environments.
%\LEm{The suggested change seemed to contradict our intended meaning. We rephrase the sentence to make it clearer.}
In conclusion, \emph{ the processes of morphological transformation,
quenching, and rapid transition to the red sequence of E galaxies seem to be in general independent
of environment}, except for a small fraction of isolated ellipticals that significantly deviate from the main
sequences of E galaxies.

From an empirical point of view, the quenching of star formation in general has been found to be associated with
mass and/or environment, mainly when the galaxy is a satellite \citep[e.g.,][]{Peng+2010}. Two main mechanisms of quenching associated with mass were proposed:
(1) the strong virial shock heating of the gas in massive haloes \citep[e.g.,][]{WhiteFrenk1991,DekelBirnboim2006},
and (2) the AGN-driven feedback acting in massive galaxies assembled by major mergers 
\citep[e.g.,][]{SilkRees1998,Binney2004}. 
Since the first models and simulations, where these mechanisms 
were implemented, it was shown that they are gradually more efficient as the halo mass increases, starting from
$\mh\sim10^{12}$ $\msun$ \citep[e.g.,][]{Granato+2004,Springel+2005,DiMatteo+2005,Croton+2006,Bower+2006,
DeLucia+2006,LCP2008,Somerville+2008}.
%IVAN: No me queda claro cual es la referencia de Dutton+2010 al respecto, asi que la saco mientras tanto. En cambio, agrego una reciente referencia de Dutton+2015 unas lineas mas abajo.
This mass corresponds roughly to %$\ms\gtrsim 3\times 10^{10}$ $\msun$.
$\ms\gtrsim 1.5 \times 10^{10}$ $\mstar$. 
As seen in Figs. \ref{gi_Ms} and \ref{sSFR_Ms},
both isolated and Coma supercluster E galaxies more massive than this follow roughly the same correlations of color and sSFR with mass, 
that is most of them are red and dead, regardless of the environment.  Therefore, \emph{for massive ellipticals, rather than the environment, 
the physical mechanisms that depend on halo mass seem to be responsible for keeping E galaxies quiescent}
\citep[see also][and more references therein]{DekelBirnboim2006,Woo+2013,Yang+2013,Dutton+2015}. 
What is the situation for the lower mass ellipticals?

As seen in Figures \ref{gi_Ms} and \ref{sSFR_Ms}, most of the E galaxies (isolated or from the Coma supercluster) with 
masses lower than %$\ms= 3\times 10^{10}$ $\msun$\
$\ms = 1.5 \times 10^{10}$ $\mstar$\ 
are also red and passive, although for these E galaxies, the
quenching mechanisms associated with mass are not already suitable. While the hostile environment of clusters 
contributes to removing and to not fostering new episodes of cold gas inflow in the E galaxies, this is not the
case for the isolated ellipticals. In fact, it should be said that most of galaxies with masses lower than 
%$\ms= 2-3\times 10^{10}$ $\msun$\
$\ms = 1 - 1.5 \times 10^{10}$ $\mstar$\ 
in the local Universe are centrals, gas-rich, blue, and SF \citep[see, e.g.,][]{Weinmann+2006,Yang+2009,Yang+2012}, 
but they are also of late type. The question for these galaxies is why they delayed their active star formation phase to a greater degree the  less massive they are \citep[this is referred to as downsizing in star formation rate; e.g.,][]{Fontanot+2009,Firmani+2010,Weinmann+2012}.
In the (rare) cases that these low-mass galaxies suffer a morphological transformation to an elliptical, 
according to our results, \emph{they should also destroy the gas reservoirs and strongly quench star formation. }

It is expected that a major merger induces an efficient process of gas exhaustion due to enhanced star formation,
but if after the merger a fraction of gas is left and/or the galaxy further accretes gas, then it could form stars again and become 
blue, SF, and even have a new disk (for theoretical works see, e.g., \citealp{Robertson+2006,Governato+2009,
Hopkins+2009,Tutukov+2011,Kannan+2015}
%IVAN: Hay dos papers Abadi+2003, pero no me queda claro si alguno de ellos corresponda a esto para major mergers. Lo saco por mientras
and for observational evidence see, e.g., \citealp{Kannappan+2009,Hammer+2009,Puech+2012}). These processes 
are very unlikely to happen in a group/cluster environment, as mentioned above, but according to our results,  late 
gas accretion and star formation do not occur  in most of the very isolated E galaxies either since they are mostly red (80\%) 
and passive (92\%), and among those that are blue (20\%), only less than one-fourth are SF. 

The quenching associated with mass can be very efficient 
for isolated E galaxies formed in haloes much more massive than $10^{12}$ $\msun$. 
In the case of ellipticals formed in less massive haloes, a possible mechanism for ejecting remaining gas or for avoiding 
further cold gas infall could be the feedback produced by type Ia supernovae (SNe Ia), which are not associated with the current star formation rate 
\citep[SFR; e.g.,][]{Ciotti+1991,Pellegrini2011}. 
Because the interstellar medium of E galaxies is very poor, the energy and momentum 
released by SNe Ia are easily expected to attain the intrahalo medium, heat it, and eventually eject it from the low-mass halo. 
Moreover, the smaller the halo, the more efficient the
feedback-driven outflows are expected to be. According to Figures
\ref{gi_Ms} and \ref{sSFR_Ms}, the lowest mass, isolated E galaxies are all red/passive, and there are pieces of evidence that the 
less massive the isolated elliptical, the faster it quenched and reddened. 
On the other hand, \citet{Peng+2015} have suggested recently that the primary mechanism responsible for quenching 
star formation is strangulation (or starvation). In this process, the supply of cold gas to the galaxy is halted with a typical timescale of 4 Gyr 
for galaxies with stellar mass less than $10^{11}$ $\msun$. However, it is not clear what could be the mechanisms behind strangulation
in an isolated environment. A possibility to have in mind is the removal of gas in early-formed, low-mass haloes due to ram pressure as
they fly across pancakes and filaments of the cosmic web \citep{Benitez-Llambay+2013}.

\subsection{Comparisons with theoretical predictions}
\label{theory}

Within the context of the $\Lambda$CDM cosmology, the hierarchical mass assembly of dark haloes 
happens by accretion and minor/major mergers \citep[see, e.g.,][and more references therein]{FakhouriMa2010}. The average
mass accretion and minor/major merger rates depend on mass and environment \citep[e.g.,][]{Maulbetsch+2007,FakhouriMa2009}. 
As a function of environment, present-day haloes in high-density regions suffered more major mergers on average and assembled 
a larger fraction of their mass in mergers than haloes in low-density regions \citep{Maulbetsch+2007}. The latter continue growing 
today mainly by mass accretion. Therefore, at face value, we expect that the very isolated galaxies could be 
efficiently accreting mass at present due to the cosmological mass accretion of their haloes.

The accretion and merger histories of CDM haloes is the first step to calculate the mass assembly and morphology 
of the galaxies formed inside them. 
In this line of reasoning, it could be expected that the isolated E galaxies formed 
in the $\Lambda$CDM cosmology should be on average significantly bluer and with higher SFRs than the `normal' E galaxies
%\LEt{Please specify what "ones" refers to.} 
formed in high-density environments because the haloes of the former continue accreting mass by today (see above). 
Nevertheless, the galaxy-halo connection is by far not direct as a result of the nonlinear dynamics of the infalling subhaloes 
and the complex physical processes of the baryons, as several semiempirical studies have shown 
\citep[see, e.g.,][]{Stewart+2009, Hopkins+2010, Zavala+2012, Avila-Reese+2014}. As the result, for instance, \citet{Zavala+2012} have shown that
%the $\Lambda$CDM scenario, despite that they implied that  halo-halo major merger rates are high, does not face a problem of  galaxy-galaxy major merger rates that are too high and the consequent overabundance of bulge-dominated galaxies. 
despite the fact that in the $\Lambda$CDM scenario the halo-halo major merger rates are high, this does not imply a problem of  galaxy-galaxy major merger rates that are too high with the consequent overabundance of bulge-dominated galaxies. 
%\LEm{The suggested change seemed to contradict our intended meaning. We rephrase the sentence to make it clearer.}

\citet{Schawinski+2009} have compared their volume-limited SDSS sample of early-type galaxies (complete to
$M_r= -20.7$ mag, which is slightly below $M^{*}$) to the $\Lambda$CDM-based semianalytical models (SAM) 
of \citet{KhochfarBurkert2005} and \citet{KhochfarSilk2006}. They found that these SAMs predict a slightly (significantly) 
higher fraction of blue (SF) early-type galaxies than the observed sample. In another work, by means of numerical 
simulations, \citet{Kaviraj+2009} have found that the expected frequency of minor merging activity at low redshift can be 
consistent with the observed low-level of recent star formation activity in some early-type galaxies \citep{Kaviraj+2007}.
However, the theoretical study that is closest to the analysis presented here is that by \citet{Niemi+2010}. 
These authors have used the SAM results of \citet{DeLucia+2007} built up in the Millennium Simulation \citep[][]{MS2005}, and applied 
some criteria to select isolated galaxies and to determine which galaxies are ellipticals. They find that $26\%$ of the 
synthetic isolated E galaxies should exhibit colors bluer than $B - R =1.4$ in an absolute magnitude range of 
$-21.5 < M_R < -20$. They call this population the \emph{blue faint isolated ellipticals.} 
In the UNAM-KIAS catalog we find that only three isolated E galaxies satisfy the previous criteria (see Fig. \ref{BR_MR}), 
which corresponds to 3.4\% of our pure E galaxy sample (0\% of ellipticals in Coma). 
These three isolated ellipticals are part of the four blue SF galaxies (blue squares in all the plots
shown in Sect. \ref{secMs}).

Part of the large discrepancy in the fractions of predicted and observed blue, faint galaxies can arise as a result of the different 
isolation criteria and morphological definitions used in \cite{Hernandez-Toledo+2010} with respect to \cite{Niemi+2010}.
The isolation criteria of the former consider neighbor galaxies as not relevant perturbers if they have a magnitude difference 
of $\Delta m_{r} \ge 2.5$ compared to an isolated galaxy candidate within a radial velocity difference $\Delta V < 1000 \textrm{ km s}^{-1}$, 
whereas in the latter the difference in magnitude is $\Delta m_{B} \ge 2.2$ inside a sphere of 500 $h^{-1}$ kpc and this condition is 
relaxed to $\Delta m_{B} \ge 0.7$ for spheres with radii between 500 $h^{-1}$ kpc and 1 $h^{-1}$ Mpc. We note that the former uses the 
$r$-band whereas the latter uses the $B$-band. Regarding the morphological definitions, in our case, galaxies are identified 
as ellipticals based on a structural and morphological analysis. They are denoted as $T \leq -4$ according to
\citet{Buta+1994}. In the case of \citet{Niemi+2010}, they use the condition $T < -2.5$ to classify modeled galaxies as ellipticals, where 
the $T$ parameter is based on the $B$-band bulge-to-disk ratio. Therefore, these authors may have included some S0 galaxies 
in their sample. 
While these differences in the isolation criteria and morphological definitions could reduce the difference
between our observed sample and the SAM prediction of \cite{Niemi+2010}, they are unlikely to explain the disagreement 
by a factor of around eight in the fractions of blue, faint isolated ellipticals.

It is known that in models and simulations the star formation of galaxies is to some point correlated with the dark matter accretion 
of their host haloes \citep[e.g.,][]{Weinmann+2012,Gonzalez-Samaniego+2014,Rodriguez-Puebla+2016}.
The fact that a substantial population of the predicted blue, faint isolated galaxies is not observed
suggests that the star formation activity of isolated E galaxies in the last Gyr(s) is overestimated in the SAMs.
The overestimate in the star formation activity is likely due to the late (dark and baryonic) high-mass 
accretion rate typical of isolated haloes (see discussion above).
Indeed, \cite{Niemi+2010} report that haloes hosting model isolated ellipticals continue their dark matter accretion 
until $z \sim 0$, whereas haloes hosting normal (nonisolated) ellipticals have joined nearly all their mass at $z \sim$ 0.5.
Furthermore, isolated ellipticals with halo masses $< 10^{12}$ $\mhalo$ have %assembled  
half of their stellar mass
by $z \sim$ 0.7, whereas isolated elliptical galaxies hosted by more massive haloes have 
half of their stellar mass 
by $z \sim$ 1.6. 
Therefore, the different mass assembly histories of the SAM isolated ellipticals hosted by low-mass haloes explain
their bluer colors compared to normal model ellipticals and other more massive isolated ellipticals.

Thus, \emph{the large difference we find here in the fraction of blue faint isolated E between observations and the
SAMs, strongly suggest that some gastrophysical processes are yet missed in the SAMs, in particular at low masses.}
We have proposed the SN Ia-driven feedback could be a mechanism that is able to avoid a
significant population of blue faint ellipticals in isolated low-mass haloes. In fact, the SAMs by \citet{DeLucia+2007} 
included the effects of the SN Ia feedback, but in a very simple (parametric) way. More detailed
studies of this process  and of the mentioned above ``cosmic web stripping'', which affects  
low-mass haloes \citep{Benitez-Llambay+2013}, are necessary.
%\LEt{Alternatively, you can divide this a sentence divided by a semicolon and say, "A more detailed study of this... is necessary; we also need to study..."\ . I did not think "namely" was necessary here.}

In spite of the differences in the fractions of blue and faint isolated ellipticals between SAMs and observations,
it should be said that the SAM predictions for the \emph{overall} population of isolated E galaxies are consistent in general
with our observed sample. The mean $B-R$ color for the synthetic isolated E galaxies in \citet{Niemi+2010} is 
1.47 $\pm$ 0.23, which is bluer than the mean for our observed isolated ellipticals (1.62 $\pm$ 0.10, see Table \ref{tabla_full}), 
but yet within the scatters.  On the other hand, the mean $B-R$ color for their modeled normal 
(nonisolated) E galaxies is 1.58 $\pm$ 0.10, which is bluer than the mean of E galaxies in the Coma supercluster (1.64 $\pm$ 0.05), 
but within the scatters again. These results show also that for both observations and models, the $B-R$ color 
distribution of E galaxies in isolated environments is broader than those in environments of higher density.

The linear fits of \cite{Niemi+2010} to their normal and isolated ellipticals are shown in the CMD of Fig. \ref{BR_MR} 
(dotted and solid lines, respectively). As can be seen, the ellipticals in the Coma supercluster follow the trend predicted by 
the model normal elliptical galaxies, but the same behavior is observed for many of our isolated ellipticals. 
\cite{Reda+2004} found a similar result using a small observational sample of six bright isolated elliptical galaxies compared to the Coma cluster.
However, there is a small fraction of isolated E galaxies that detach from this trend, toward bluer colors at fainter magnitudes. 
This trend is similar to that followed in the SAM by the isolated ellipticals, while, as reported above,
the fraction of galaxies following this trend is much higher in the SAM than in observations.  
In the ($B-R$)--\mh\ diagram (left panel of Fig. \ref{BR_Mh}), we plot with a dotted line a visual approximation for the trend 
of the model nonisolated ellipticals in \cite{Niemi+2010}. These authors predict that normal ellipticals have a roughly constant color 
for the whole halo mass range sampled ($10^{10} < M_h/\mhalo < 10^{14}$), but isolated ellipticals show bluer colors at $M_h < 10^{12}$ $\mhalo$. This behavior is somewhat similar for our observed isolated E galaxies, thus suggesting that some isolated ellipticals hosted by low/intermedium-mass haloes should exhibit bluer colors than normal ellipticals in haloes with the same mass.

We conclude that the $\Lambda$CDM-based models of galaxy evolution are roughly consistent with observations in 
regards to the local population of E galaxies, both isolated and in cluster-like environments. However, \emph{the SAMs predict
a too high abundance of blue, faint isolated  ellipticals formed in low-mass haloes with respect to our observational inference.}
The above discussed mechanisms (SN-Ia feedback and cosmic web stripping) could help to avoid further gas accretion to 
galaxies in low/intermedium-mass haloes.
%but in this case the question would be now why
%there is a small fraction of isolated ellipticals that are blue and SF. In order to answer to this question, we 
%need to analyze in more detail the properties of these galaxies. This is the aim of the next section.

%============================================================
\section{The blue SF isolated ellipticals} 
\label{blueSF_Es}
%============================================================

In the previous Sections, we have shown that isolated E galaxies, as well as those in the Coma supercluster, are mostly
red and passive. However, in the isolated environment, there is a small fraction of ellipticals that systematically
deviate from the red/passive sequence as their masses are smaller. %their masses. 
The question is whether these few 
intermedium-mass ellipticals are blue and SF because they suffered  the morphological transition recently 
and did not yet exhaust their gas reservoir \citep{McIntosh+2014,Haines+2015}, which can include the process of
disk regeneration  as suggested by \citet[][]{Kannappan+2009}, or because these ellipticals were rejuvenated 
by recent events of cold gas accretion as suggested by \citet[][]{Thomas+2010}.

In general, several pieces of evidence show that the structural properties and correlations of blue SF 
isolated ellipticals do not differ significantly from  those of the other isolated ellipticals or even the group/cluster
ellipticals. For example,  in Fig. \ref{R50_Ms} we show that the radius--\ms\ correlation of all ellipticals is roughly the same.
 \citet{Kannappan+2009} found that blue-sequence E/S0s are more similar to red-sequence E/S0s 
than to late-type galaxies in the \ms--radius relation. Blue E/S0 galaxies are also closer to red E/S0 than 
to late-type systems in this relation at 0.2 $< z <$ 1.4 \citep{Huertas-Company+2010}.
We need to go in more detail and explore whether the blue SF isolated ellipticals have some
peculiarities that could suggest us which mechanisms are dominant in making them blue and SF.

A characterization of first order on the nature of the blue SF isolated E galaxies is 
proposed here  using the available $gri$ images and spectra from the SDSS database. 
This includes only four galaxies, three of which coincide with the definition of blue, faint isolated 
ellipticals in \citet{Niemi+2010}. 
Table \ref{tabla_general} summarizes the general properties of these galaxies. 
In Apprendix \ref{Ap_BlueSF}, we present an analysis of several structural-morphological and 
spectroscopical properties for each one of the four observed blue SF isolated ellipticals. 
We point out that UNAM-KIAS 1197 shows evidence that it is  classified as a LINER galaxy
and that UNAM-KIAS 613 could be actually an AGN due to its broad component in H${\alpha}$
(see Appendix \ref{Ap_BlueSF} for details).
We have also carried out similar analyses for the other, more common red and passive 
isolated ellipticals (to be presented elsewhere).
In the following, we discuss the results of our analysis with the aim of elucidating  the nature of the blue SF isolated ellipticals.

%%% TABLA GenProp
\begin{table*}  
\caption{General properties of the blue SF isolated elliptical galaxies. 
%classified as blue faint, red SF and red passive galaxies.
}
\centering
\begin{tabular}{c c c c c c c c c c}
\hline
\hline
  name         & $z$&$\log(\ms)$& $g-i$ &log(sSFR)  & $^{0.0}M_{\rm R}$ & $B - R$ & R$_{deV}$ & age & $\log(\mh)$\\
%               &    &($\mstar$)&       &(yr$^{-1}$)&                   &         &        & (9) & (10) \\
\hline
%      Blue Faint Ellipticals \\
UNAM-KIAS 359  &0.017& 9.86    & 0.94  &-10.26   &  -20.50           & 1.38    &  1.50     &868.0&     ... \\
UNAM-KIAS 613  &0.027& 10.30   & 1.04  &-10.25   &  -21.23           & 1.48    &  1.53     &372.5& 11.93\\
UNAM-KIAS 1197 &0.032& 10.15   & 0.77  &-10.50   &  -21.38           & 1.21    &  2.13     &546.8& 11.78\\
UNAM-KIAS 1394 &0.035& 10.13   & 0.81  & -9.63   &  -21.34           & 1.25    &  1.01     &268.2& 11.75\\
\hline
\end{tabular}

\tablefoot{Columns are: name in the UNAM-KIAS catalog, redshift,
%physical scale (kpc, $h = 0.73$) subtended by the 3 arcsec SDSS fiber at the distance of each galaxy. 
log$_{10}$ of the stellar mass in units of $\mstar$, $g-i$ color,
log$_{10}$ of the specific star formation rate in units of yr$^{-1}$,
absolute magnitude in the $R$-band ($h=0.7$), $B-R$ color,
radius of a de Vaucouleurs fit in the $r$-band in units of kpc ($h=0.7$), 
luminosity-weighted stellar age in units of Myr from the STARLIGHT database, 
and, if it is available, log$_{10}$ of the host halo mass from Y07 in units of $\mhalo$.
}
%\LEm{Units for halo mass are $\mhalo$. The halo mass is not available for UNAM-KIAS 359.}
\label{tabla_general}
\end{table*}

From the surface brightness profiles in the bands $g$ and $i$, we find that the four blue SF isolated ellipticals 
show radial color gradients with bluer colors toward the galaxy center (see bottom- left panels in 
Figs. \ref{mosaico359}, \ref{mosaico613}, \ref{mosaico1197} and \ref{mosaico1394} in the Appendix \ref{Ap_BlueSF}),
while most of the red and passive ellipticals show negative or flat radial color profiles 
\citep[e.g.,][]{denBrok+2011}.
The positive color gradient may be evidence of dissipative infall of cold gas, which promotes recent 
star formation in the central regions. This supports the rejuvenation scenario for the blue SF ellipticals.  
\citet{Suh+2010} have suggested that 
positive color gradients in early-type galaxies are visible only for
$0.5 - 1.3$ billion years after a star formation event. Afterward, the galaxies exhibit negative color gradients.
\citet{Shapiro+2010} have also proposed the rejuvenated scenario for some early-type galaxies in the SAURON sample \citep{deZeeuw+2002}  %\citep{Bacon+2001} 
with red optical colors that show star formation activity in the infrared. These galaxies correspond to fast-rotating systems with concentrated star formation.
They suggest that when the star formation ceases, over the course of $\sim 1$ Gyr, the transiently star-forming galaxy return to evolve passively. 
Furthermore, \citet{Young+2014} have also proposed the rejuvenation scenario to explain the blue tail of early-type galaxies in the color--magnitude diagram in the
ATLAS$^{\verb|3D|}$ sample \citep{Cappellari+2011}.

From our detailed photometric analysis (Appendix \ref{Ap_Phot}), we find that the structure of isolated E galaxies can be described in general by three 
S\'ersic components (inner, intermediate, and outer), which 
is usually reported for ellipticals in groups and clusters \citep[e.g.,][]{Huang+2013}.
However, the outer components of our blue SF isolated ellipticals (see top-right panels of Figs. \ref{mosaico359} \ref{mosaico613}, \ref{mosaico1197} and \ref{mosaico1394} and Table \ref{tabla_Phot} in Appendix \ref{Ap_BlueSF}) are present in two cases, UNAM-KIAS 613 and UNAM-KIAS 1197, 
where the latter shows  a small value of $n$, and in other two cases it seems that this component is even absent, 
UNAM-KIAS 359 and UNAM-KIAS 1394. 
On the contrary, bright cluster galaxies have been reported to have extended stellar envelopes with large $n$ indexes 
\citep{MorganLesh1965,Oemler1974,Schombert1986}. 
For a sample of low-luminosity ellipticals, most of them in group-like environments, \citet{Huang+2013} measured a mean $n$ value of 1.6 $\pm$ 0.5 and a mean effective radius $r_{eff} = 7.4 \pm 2.6$ kpc for the outer
component. Only UNAM-KIAS 613 has a $n$
value larger than this mean and all the blue SF isolated ellipticals have values of the outer $r_{eff}$ that are smaller than the
mentioned mean. Therefore, at least three of the four blue SF isolated ellipticals seem to have an outer structure that is different 
from other ellipticals.

According to \citet{Huang+2013_L28}, the different structural components in E galaxies may be explained by 
a two-phase scenario \citep{Oser+2010,Johansson+2012}.  The inner and/or intermediate components are the outcomes of an initial 
phase characterized by dissipative (in situ) processes such as cold accretion or early gas-rich mergers. The outer, extended component 
of ellipticals are related to a second phase dominated by nondissipative (ex situ) processes such as dry, minor mergers after the quenching 
of the galaxy.  This could also explain the build up of E galaxies in isolated environments. Galaxies with very small outer $n$ and $r_{eff}$ 
values or with no outer component, which is the case in three out of four blue SF isolated ellipticals, could have not suffered ex situ 
processes recently. This again supports the rejuvenation scenario for the blue SF ellipticals, at least for three of them.

Regarding the inner components of the blue SF isolated ellipticals, the results of our study show that their best-fit S\'ersic indices have
$n \leq 2.0$ and $r_{eff} \leq 0.6$ kpc (see Table \ref{tabla_Phot}  in Appendix \ref{Ap_BlueSF}), whereas \citet{Huang+2013} obtained 
mean values of $n = 3.2 \pm 2.1 $ and $r_{eff}$ = 0.7 $\pm$ 0.4 kpc for their 
%(low-luminosity) 
sample of ellipticals.  Thus, though
within the scatter, the blue SF isolated ellipticals seem to have a more disky inner component than other low-luminosity ellipticals. 
This can be consistent with both the rejuvenation by cold gas infall or the post-merger disk regeneration scenarios.

In Appendix \ref{fine-structure}, we list the different fine structure and residual features that 
can be found in the images of E galaxies and their association with different levels of interaction/merger.
Our analysis (see Table \ref{tabla_Phot}) shows that two of the blue SF isolated ellipticals
(UNAM-KIAS 613 and UNAM-KIAS 1394) do not present convincing evidence of (recent) disturbances of any type
and the other two, UNAM-KIAS 1197 and  UNAM-KIAS 359, present evidence of weak disturbance effects
through low surface brightness (LSB) outer shells.
We found no evidence of tidal tails or broad fans of stellar light, which are both associated with dynamically cold components produced
by an accreted major companion. No evidence of significant sloshing ($>10\%$) in the inner kpc region was found either. 
An additional revision of the residuals in the very central regions suggests the presence of a few thin localized patches, which 
we tentatively interpret as coming from dusty features. This is consistent with the reddened central subregions observed in the 
corresponding color maps. Such presumably nuclear dust structures may be associated with the inner or intermediate 
disk-like components found in our decomposition analysis, and could be evidence of centralized star formation due to cold 
gas infall. Thus, the lack of evident fine structure and residual features and the presence of nuclear dust structures
in the four blue SF isolated ellipticals supports again the mechanism of rejuvenation in contraposition to the one
of recent mergers and disk regeneration. 
We caution, as noted by \citet{GZ2015}, that the SDSS images are not the most adequate for detecting finer details in early-type galaxies. Deeper imaging data is preferable.

Finally, in the Appendix \ref{Ap_Spec} we describe our spectroscopic analysis for the four blue SF 
isolated elliptical galaxies, where the SDSS spectra were used. Recall that SDSS provides only one optical fiber of 3 arcsec 
aperture centered in each galaxy. For our four objects, this corresponds to the inner $\approx 1-2$ kpc, which roughly corresponds
to their de Vacouleours effective radii (Fig. \ref{R50_Ms}). Our analysis of the emission lines gives similar results
in the BPT diagram as those reported in the STARLIGHT database and plotted in Fig. \ref{BPT} above (the loci of our 
estimates are indicated with a blue open square). 
We note that UNAM-KIAS 613 has very broad components in H${\alpha}$ 
 and a clear pseudocontinuum (see Figs. \ref{mosaico613} and \ref{broad613} in Appendix \ref{Ap_BlueSF}). UNAM-KIAS 1197
qualifies as a LINER, UNAM-KIAS 359 as a TO, and only UNAM-KIAS 1394 shows a clear evidence of dominant 
central star formation.
Our analysis suggests then that the ionization mechanism could have a large contribution from a nuclear nonthermal 
component (e.g., LINER) in two of the four blue SF isolated ellipticals. Thus, the star formation rate in these
galaxies could be lower than that calculated from the H${\alpha}$ flux. 
%HECTOR: Ojo con esta conclusion y con la que sigue basada en luminosity-weighted average ages!!!

The stellar populations encode the mass assembly history over the lifetime of the blue SF isolated elliptical galaxies, 
which is important to gain insights on their formation and evolution. 
By applying a stellar population synthesis analysis (see Appendix \ref{Ap_Spec}),
the obtained mass-weighted star formation histories show that the four blue SF ellipticals formed $<5\%$ ($<20\%$) %at
out of
their present-day stellar masses in the last 1 (3) Gyrs (see Fig. \ref{age_STAR} and Table \ref{tabla_Spec}).
The obtained light-weighted star formation histories 
show that 30--60\% of the present-day luminosity is due to star formation in the last 1 Gyr.  
These results suggest that
the blue SF isolated ellipticals formed most of their stars early, 
%have had a period of enhanced star formation in the last $\sim 1$ Gyr. 
but in the last $\sim 1$ Gyr they had a period of enhanced star formation.
This enhanced period of star formation is reflected in the very small luminosity-weighted average ages obtained
for these galaxies with respect to the rest of the ellipticals 
(see Fig. \ref{ageF_Ms}).
%HECTOR: Solo que en esta conclusion al estar considerando luminosity-weighted average ages, estas 
%tomando en cuenta la contribucion del AGN tambien!!!! 

We can also estimate the star formation timescale \citep[SFTS;][]{Plauchu-Frayn+2012} activity in each galaxy by calculating the difference of mass-weighted and light-weighted average stellar ages  (age$_{mw}$ and age$_{lw}$, respectively) as  $\Delta$(age) = 10$^{\rm{log(age_{mw})} }$ - 10$^{\rm{log(age_{lw})} }$. These values are reported in Table \ref{tabla_Spec}.
The SFTS is an indicator of how fast a galaxy created its stellar population, or how long the stellar activity was prolonged in this galaxy. A typical elliptical galaxy, for example, where star formation has stopped long ago, would be expected to have a short SFTS. 
The blue SF isolated E galaxies have SFTS values of 8.5 Gyr on average. As a comparison, \citet[][]{Plauchu-Frayn+2012} find that early-type galaxies in Hickson compact groups \citep[][]{Hickson1982,Bitsakis+2010,Bitsakis+2015} have SFTS values of 3.3 Gyr, whereas similar isolated galaxies have values of 5.4 Gyr, meaning that the former have formed their stars over shorter timescales than isolated early-type galaxies. The blue SF isolated galaxies show higher values than those respective samples of early-type galaxies, which suggests a more prolongated star formation activity in these galaxies.

Our photometric and spectroscopic analyses of the rare four blue SF isolated ellipticals %is 
are not conclusive. 
However, our %analysis
results suggest that in general these galaxies do not present 
evidence of strong recent disturbances or mergers in their
structure and morphology but they have been forming stars since the last $\sim 1$ Gyr in the central regions; 
in two cases there is  also some evidence of AGNs. We conclude that \emph{it is more plausible that these isolated E galaxies
assembled early as other ellipticals but they were rejuvenated by recent ($<1$ Gyr)  accretion events of 
cold gas.}
On the other hand, integral field spectroscopy (IFS) has allowed %us to
the study of kinematic and stellar population properties of early-type galaxies. For example, in the SAURON \citep{Bacon+2001,Kuntschner+2010,Shapiro+2010}, ATLAS$^{\verb|3D|}$
\citep{Cappellari+2011,Young+2014,McDermid+2015}, and CALIFA \citep{Sanchez+2012,GonzalezDelgado+2014,GonzalezDelgado+2015} projects.
Future observations with IFS
will be crucial to shed more light on the formation and evolution 
%of this particular class 
of blue isolated elliptical galaxies.

In the Introduction we stated the possibility of using pure E galaxies in very isolated environments as `sensors' of 
gas cooling from the intergalactic medium. This cool gas, once trapped by a galaxy, should form stars.
The fact that only a negligible fraction ($\lesssim 4\%$) of our sample of local isolated ellipticals
are blue and SF suggests that the process of cooling and infall of gas from the warm-hot intergalactic medium 
is very inefficient.

%========================
\section{Conclusions} 
\label{Conclusions}
%========================

We have studied a sample of 89 local very isolated E galaxies ($z=0.037$ on average) and compared their properties 
with those from E galaxies located in a higher density environment, the Coma supercluster. The samples studied here refer only to morphologically, well-defined elliptical (pure-spheroid) galaxies %(T $\leq -4$) 
in the mass range 
%$10^9\lesssim \ms/\msun \lesssim 4\times 10^{11}$, 
$6 \times 10^{8}\lesssim \ms/\mstar \lesssim 2\times 10^{11}$, 
in contrast to other works that select overall early-type galaxies 
including S0s objects.  Our main results and conclusions are as follow.

(i) The correlations of color, sSFR, and size with mass that follow most of the isolated E galaxies are similar to those of the Coma supercluster 
E galaxies. Notwithstanding the environment, most of ellipticals are `red and dead'.
%``red, dead, and compact''. 
All E galaxies more massive than 
%$\ms\approx 10^{11}$ $\msun$ ($\mh\approx 10^{13}$ $\msun$) 
$\ms\approx 5 \times 10^{10}$ $\mstar$ ($\mh\approx 10^{13}$ $\mhalo$)
are %passive. 
quiescent.
As the mass or luminosity is smaller, both isolated and Coma 
ellipticals become bluer, although on average they remain in the red sequence. However, a few intermediate-mass ellipticals
pass to be moderately blue in Coma, while in the  case of the isolated ellipticals, a fraction of them deviates 
systematically toward the blue cloud. The extreme of this branch is traced by those isolated ellipticals that are blue and SF at 
the same time; this includes only four ellipticals, which  have intermediate stellar masses between 
%$1.4\times 10^{10}$ and  $4\times 10^{10}$ $\msun$.
$7\times 10^{9}$ and $2\times 10^{10}$ $\mstar$.  
These blue SF isolated ellipticals are also the youngest galaxies with light-weighted stellar ages $\lesssim 1$ Gyr,
which could indicate recent processes of star formation in them. 

(ii) In terms of fractions, among the isolated ellipticals $\approx 20\%$ are blue, $\lesssim 7\%$ are SF, and $\approx 4.5\%$ are 
blue SF, while among the Coma ellipticals $\approx 8\%$ are blue, $\la 1 \%$ are SF, and there are no blue SF objects.  On 
average, the galaxies in Coma have sSFR values that are lower than isolated ellipticals by $\sim 0.2$ dex and are older by $\la 1$ Gyr.
Based on a color--color criterion, $\approx 10\%$ of the isolated ellipticals show evidence of recent quenching. All of these isolated 
RQEs are less massive than 
%$\ms \approx 1.5\times 10^{11}$ $\msun$,
$\ms \approx 7 \times 10^{10}$ $\mstar$, 
and are approaching the red sequence (the two lowest massive
ellipticals  are actually already red), which suggests that the quenching and reddening happened quickly in the isolated environment.  
In the Coma supercluster, excepting one marginal case, there are no RQEs, even among those that are still blue;
%\LEt{"yet blue" is ambiguous. How about, "...those that are not yet blue/yet to become blue."?} 
for the latter, the 
quenching and reddening seem to have proceeded more gradually.

(iii) Around $40\%$  %$38-39\%$ 
of the E galaxies have detectable (S/N $>3$) emission lines in both isolated and dense environments. 
According to the BPT diagram, most of these are AGNs. However, the fraction of those classified as SFN/TO is slightly
larger for the isolated ellipticals than for those in the Coma supercluster (11\% and 8\%, respectively).

Our results show that all massive ellipticals 
%($\ms\gtrsim 10^{11}$ $\msun$),
($\ms\gtrsim 5 \times 10^{10}$ $\mstar$), 
and a large fraction of the less massive ellipticals, assembled 
their stellar populations early, remaining quenched since these epochs, regardless of whether they are isolated or from the Coma supercluster.  
Moreover,  in both of these different environments, a downsizing trend is observed: as the mass becomes lower, the ellipticals are on average less red and 
have higher sSFR. Thus, rather than environment, it seems that the processes involved in the morphological transformation of 
E galaxies are those that dominate in their efficient star formation shut-off, the depletion of their cold gas reservoir, and their downsizing trends. 
On the other hand, new episodes of cold gas inflow are very unlikely to happen in the environment of clusters or for isolated galaxies
living in massive haloes, hence, the E galaxies are expected to remain quenched. However, our study shows that most of intermediate- 
and low-mass isolated ellipticals have also transited to the red/passive sequence by $z \sim 0$. We suggested two possible mechanisms
to explain why most low-mass E galaxies in an isolated environment could be devoid of gas: (1) the galactic winds produced by
the feedback of SNe Ia, and (2) the removal of gas in low-mass haloes due to ram pressure as they fly across pancakes and 
filaments of the cosmic web. 
  
Interesting enough, the predictions of $\Lambda$CDM--based SAMs for the population of E galaxies (both in clusters/groups and isolated)
agree in general with the results of our study, except that these models predict a too high abundance (a factor of 8 more) 
of blue, faint (low-mass) isolated ellipticals with respect to our results. This suggests that some gastrophysical processes at 
low masses, for example, those
mentioned above, are yet missed or underestimated in SAMs.

Our results show that E galaxies in the isolated environment are not too different from those in the Coma supercluster at a given mass,
but the fractions of blue or SF objects is larger in the former case than in the latter. Hence, in some cases, the isolated environment seems to 
propitiate  the rejuvenation  or a late formation of the ellipticals. The extreme examples are those ellipticals that 
are blue and SF at the same time; they exist only in the isolated environment.  In Appendix \ref{Ap_BlueSF}, we 
presented a structural/spectroscopic analysis of these four ellipticals with the aim of inquiring about their nature. We  found the following
for them:
%with stellar masses of $M_{s} = 10^{9.8} - 10^{10.3}$ $\mstar$. 

(iv) The four blue SF isolated ellipticals have radial color gradients with bluer colors toward the galaxy center.  
Furthermore, at least three out of the four blue SF isolated ellipticals have only two (inner/intermediate) structural 
components, lacking the third outer component seen in classical ellipticals. The four ellipticals lack significant fine 
structure and residual features, and show the presence of nuclear dust structures. 

(v) The spectroscopic analysis suggests that the ionization mechanism can have a large
contribution from a nuclear nonthermal component (e.g., LINER) in two of the four blue SF isolated ellipticals. 
On the other hand, $30 - 60\%$ of their present-day luminosity, but only $<5\%$ of their present-day mass, 
is due to star formation in the last 1 Gyr. This suggests that these galaxies formed most of their stars early
but in the last $\sim 1$ Gyr they had a period of enhanced star formation. 
Their high SFTS values suggest that they have formed their stars over prolongated timescales.

The positive color gradient in the four blue SF isolated ellipticals may be evidence of recent cold gas infall, 
which supports the rejuvenation scenario in contraposition to the scenario of a recent merger and/or disk regeneration.
The presence of only inner and intermediate structural components, which are related to dissipative processes such 
as cold accretion or early gas-rich mergers, and the lack of  outer component, which is related 
to nondissipative processes (e.g., dry mergers) after the quenching of the galaxy, suggest that these ellipticals
did not suffer recent dry mergers but probably had cold gas accretion. This is supported by the lack of fine structure and residual
features and the nuclear dust structures. 

We conclude that it is more plausible that the blue SF isolated E galaxies assembled early as other ellipticals, 
but they were rejuvenated by recent ($< 1$ Gyr) accretion events of cold gas. Further work with powerful observational 
methods such as IFS is needed to investigate the kinematic and stellar population properties resolved in space of 
the blue SF isolated elliptical galaxies. These galaxies can be used to trace and estimate the fraction of recent gas 
cooling from the cosmic web.

\begin{acknowledgements}

We thank Mariana Cano-D\'{i}az for her valuable comments and suggestions about the modeling of the UNAM-KIAS 613 spectrum. We acknowledge the anonymous referee for his/her constructive comments on our paper. 
HMHT and VAR acknowledge CONACyT grant (Ciencia B\'asica) 167332 for partial support. 
%JAS ...
HMHT and JAS acknowledge DGAPA PAPIIT IN-112912 for financial support.
ADO acknowledges financial support from the Spanish Ministry for Economy and Competitiveness through grant AYA2013-42227-P and from Junta de Andaluc\'{i}a TIC114.
This research has made use of the GOLDMine Database.
%operated by the Universita' degli Studi di Milano- Bicocca.
This research has also made use of the NASA/IPAC Extragalactic Database (NED) which is operated by the Jet Propulsion Laboratory, California Institute of Technology, under contract with the National Aeronautics and Space Administration.

\end{acknowledgements}

\bibliographystyle{aa}
\bibliography{references}
%\end{document} 

\begin{appendix}

\section{Results for individual blue SF isolated elliptical galaxies}
\label{Ap_BlueSF} 

Various image procedures were applied to the SDSS $gri$-band images to extract information
about the internal and external structure of these galaxies. The details of the methodology
are in Appendix \ref{Ap_Phot}. On the other hand, a population synthesis model was 
applied to the SDSS spectra to extract information on the stellar populations and gas 
properties of their central regions. The methodology is explained in Appendix \ref{Ap_Spec}. 
The results of each galaxy are summarized in Tables \ref{tabla_Phot} and \ref{tabla_Spec} and
presented in Figs. \ref{mosaico359} , \ref{mosaico613}, \ref{mosaico1197}, and \ref{mosaico1394} with the following set of panels: 
%\begin{item}
\begin{itemize}
\item (a) Top left. First row from left to right shows an $gri$ image of the galaxy from the SDSS database, a sharp-filtered image with a $\sigma$ = 4 for the Gaussian function (in pixels) along the direction $\theta$ of the major axis of the Gaussian function and $g - i$ color index map. %(first row from left to right in the top-left panels).
Second row from left to right shows
the best set of S\'ersic profiles for the observed surface brightness 
distribution in the $r$-band, which is summarized as the residual image after subtracting the best first model (S\'ersic1), %the residual image after subtraction of 
the best second model (S\'ersic2), and the best third model (S\'ersic3). 
Images are registered at the nuclear region of each galaxy. The individual components found in the 2D decomposition are 
roughly represented with ellipses in different colors (red, blue, green, and cyan); 
each ellipse shows the semimajor axis in units of $r_{eff}$ 
using the $b/a$ axis ratio of each best S\'ersic model. 

\item (b) Top right. A comparison of azimuthally averaged 1D $r$-band profiles 
of the data and the best global model along with the corresponding residuals (lower box). Both the observed and best-model azimuthally averaged 1D profiles were extracted by imposing a fixed center and free ellipticity $\epsilon$ and position angle estimates. 
In general, the azimuthally averaged 1D surface brightness profiles built from this model solution reproduce  the observed profiles reasonably well.

\item (c) Bottom left. Surface brightness profile in the $g$ and $i$ bands. The profiles were extracted by imposing a fixed center and fixed ellipticity $\epsilon$ and position angle estimates to guarantee homogeneous color estimates.
The lower panel shows the $g-i$ color gradient, corrected for  galactic extinction.  

\item (d) Bottom right. The observed and modeled spectrum with STARLIGHT, and the residual spectrum (lower box).

\end{itemize}

In the following analysis, we use $h = 0.73$, $\Omega_M=0.27$, $\Omega_{\Lambda}=0.73$
along with a correction by Virgo centric infall to estimate the physical scales (e.g., $r_{eff}$ 
and fiber size).
We here resume the main features found from our analysis for each  of the four blue SF isolated ellipticals. We 
explicitly repeated the spectroscopic analysis for the following four isolated ellipticals by applying the available STARLIGHT code, 
modeling their one-fiber SDSS spectra, and then fitting the emission lines %remaining in 
from the residual (observed - modeled) spectra.

\subsection{UNAM-KIAS 359}       

\begin{figure*}
\centering
\includegraphics[width=15cm]{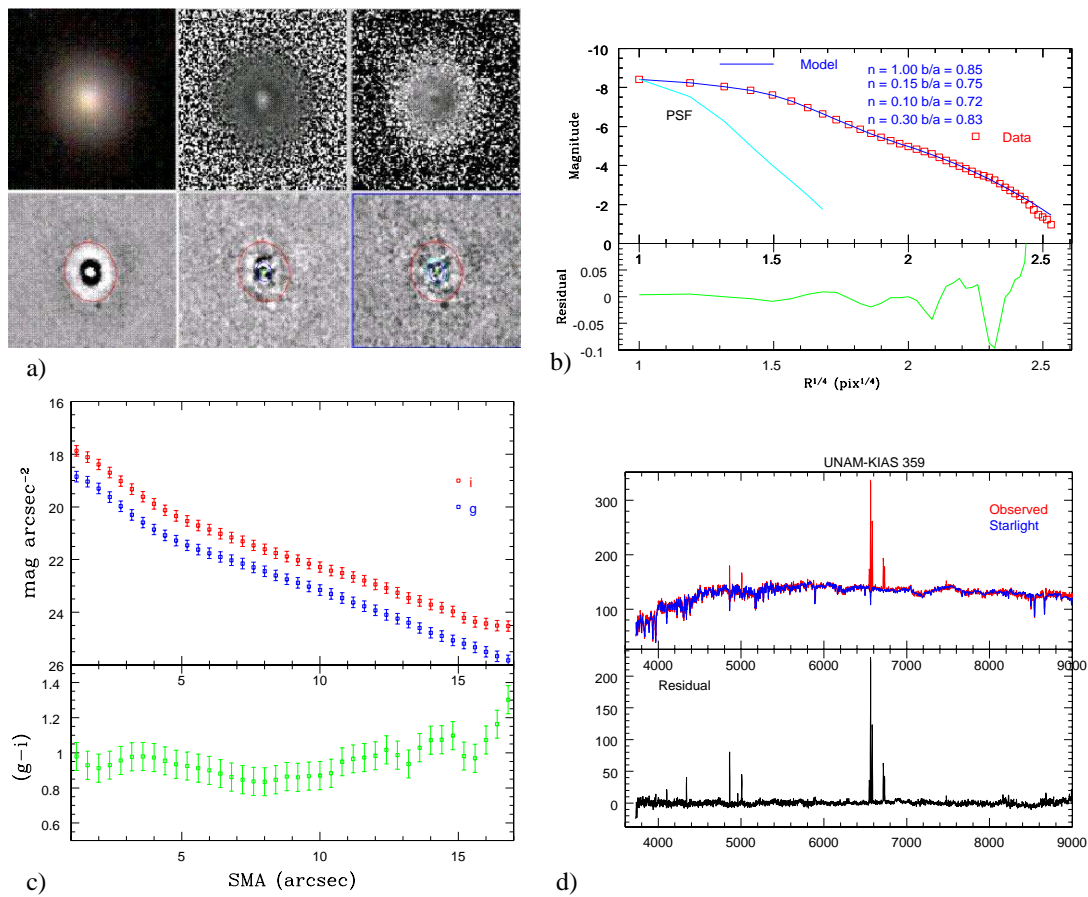}
\caption{
Analysis for the blue SF galaxy UNAM-KIAS 359. Top-left panels (a): first row shows (from left to right) the SDSS $gri$-band color image, filter-enhanced $r$-band image, and $g-i$ color map.
Second row shows (from left to right) residual $r$-band images from 
S\'ersic1, S\'ersic2, and S\'ersic3 models. 
The red ellipse shows the geometric parameters for one Sersic component, whereas additional components are shown in blue, green, and cyan ellipses if they correspond.
Top-right panels (b): azimuthally averaged surface brightness distribution (red squares) and that derived from the best S\'ersic components model (blue line). PSF modeling of field stars is shown as a cyan line. The residual comparison is shown in the lower box (green line). 
Bottom-left panels (c): surface brightness profile in the bands $g$ and $i$ (blue and red squares, respectively). The $g-i$ color gradient is shown in the lower box.
Bottom-right panels (d): observed SDSS spectrum (red line) along with the corresponding modeled population synthesis spectrum (blue line). The residual spectrum is shown in the lower box (black line).
}
\label{mosaico359}
\end{figure*}

No similar-sized neighbor galaxy is found within %35 arcmin 
727 kpc
and $\pm$1000 km s$^{-1}$. The color map in the top-left
panel of Fig. \ref{mosaico359} shows a conspicuous central region with a uniform reddened cusp but asymmetric 
reddening distribution, an intermediate bluer region, and an outer diffuse red region. 
%with red. 

Our 2D residual images show %(i) 
some tiny central residuals consistent with the color index map showing a strong 
reddened central cusp and an adjacent reddened patch to the right, probably associated with dusty structures. There is
marginal evidence of external low surface brightness features. This galaxy does not show any significant sloshing in the 
inner kpc region. This galaxy has been reported as detected in \ion{H}{i} 21 cm. 

Our analysis from GALFIT 
shows %four components (top-right panels); 
three inner components ($r_{eff}$ = 0.15, 0.55, and 
0.80 kpc), all with S\'ersic indices lower than 1 ($n = 0.1$, 0.15, and 0.30, respectively), and one intermediate-outer component
($r_{eff}$ = 1.95 kpc) with $n = 1.00$.

At the distance of UNAM-KIAS 359, the 3 arcsec SDSS fiber spectrum subtends 1.04 kpc. 
From our spectroscopic analysis, this galaxy qualifies in the BPT diagram as a transition object (TO) with a mixture of a 
central, low-luminosity AGN and emission coming from a circumnuclear SF region (see Fig. \ref{BPT}).

%%% FigA1

\subsection{UNAM-KIAS 613}

\begin{figure*}
\centering
\includegraphics[width=15cm]{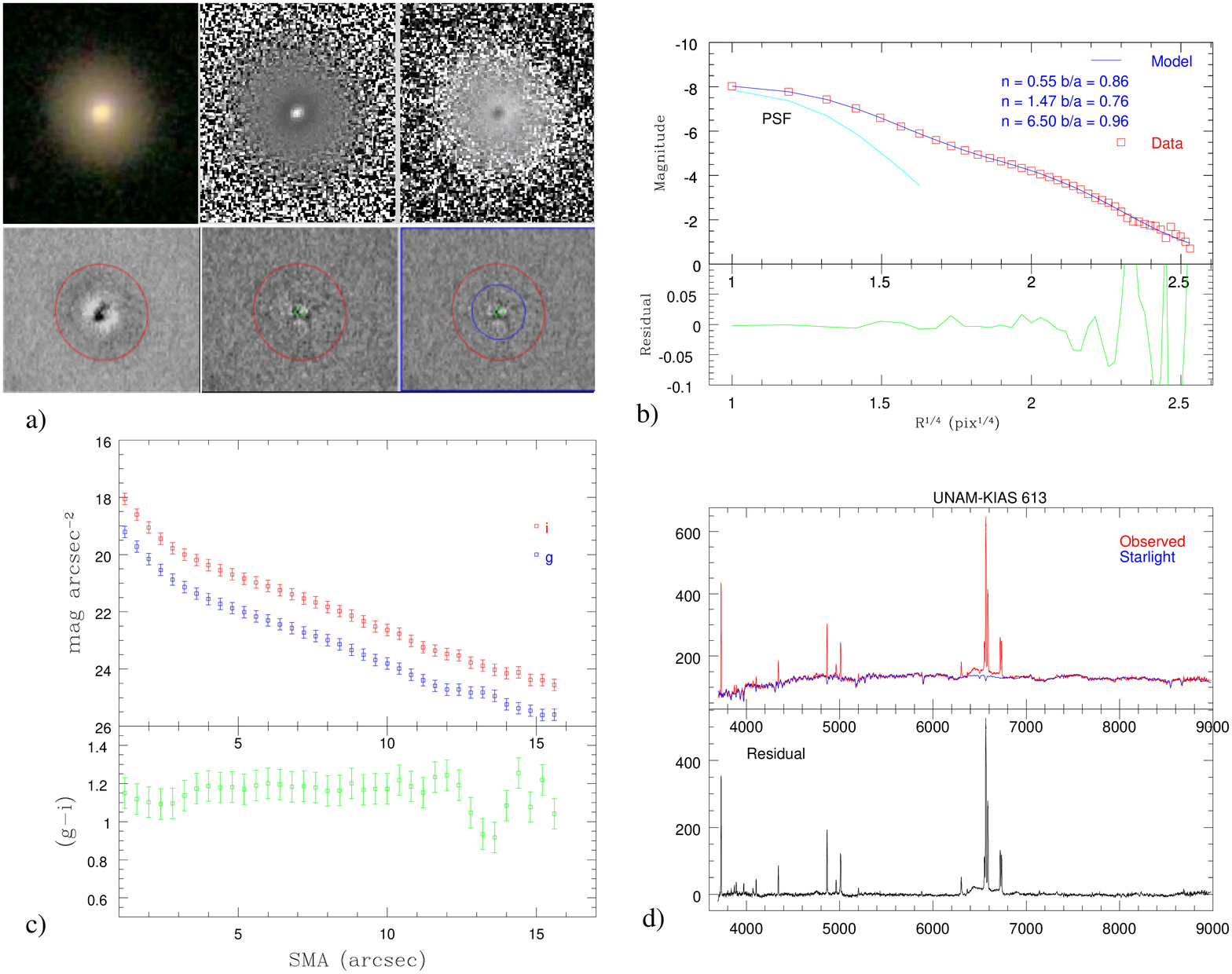}
\caption{
Similar to Fig. \ref{mosaico359}, but for UNAM-KIAS 613.
}
\label{mosaico613}
\end{figure*}

Only one neighbor galaxy fainter than 2.5 mag with respect to UNAM-KIAS 613 is within %22 arcmin
712 kpc
and $\pm$1000 km s$^{-1}$.
The SDSS images in the top-left panel of Fig. \ref{mosaico613} show (i) 
a conspicuous nuclear region (maybe a lens) surrounded by (ii) an adjacent more diffuse region and also an (iii) outer 
envelope. The central strongly reddened cusp region contrasts with the more bluish colors of the rest of the galaxy in 
the color map.  

The 2D residual images show a few weak filamentary features, probably associated with dust structures. The sloshing is not 
significant in the inner kpc. No low surface brightness features are observed in the outer regions of this galaxy.
The galaxy was classified as S0 by \cite{NA2010}. 

Our analysis from GALFIT, summarized in the top-right panel of Fig. \ref{mosaico613}, shows three components; 
an inner one ($r_{eff}$ = 0.43 kpc) with S\'ersic index  $n = 1.47$, an intermediate one ($r_{eff}$ = 2.81 kpc) with 
$n = 0.55,$ and an outer one ($r_{eff}$ = 4.54 kpc) with $n = 6.50$. 

\begin{figure*}
\centering
\includegraphics[width=15cm]{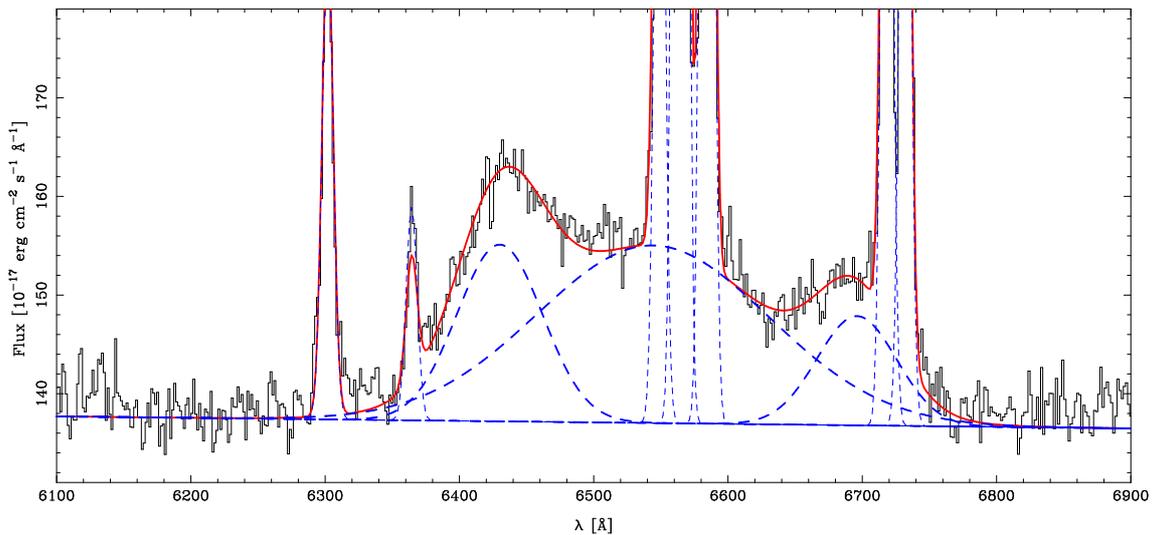}
\caption{
Spectrum of the H${\alpha}$ region of UNAM-KIAS 613. Horizontal axis is in the rest-frame wavelength. Black line corresponds to the original host galaxy subtracted spectrum, while the red line shows the model including all the emission-line components. The blue thick dashed lines trace the broad central component and the blue and red broad components of the H${\alpha}$ line, while the blue thin dashed lines show the individual profiles associated with the narrow components (H${\alpha}$, [\ion{O}{i}]6300,6364, [\ion{N}{ii}]$6548,6584$, [\ion{S}{ii}]$6717,6731$).
}
\label{broad613}
\end{figure*}

At the distance of UNAM-KIAS 613, the 3 arcsec fiber from the SDSS spectrum subtends about 1.62 kpc.
From our spectroscopic analysis, this galaxy initially qualifies as a TO (see Fig. \ref{BPT}) with broad emission lines (see panel d in  Fig. \ref{mosaico613}). 
UNAM-KIAS 613 is the only galaxy in the sample where a broad component in H${\alpha}$ is clearly detected as can be seen in more detail in Fig. \ref{broad613}. Also a fainter broad component is detected in H${\beta}$ (see panel d in Fig. \ref{mosaico613} where the whole spectrum is plotted) and consequently, according to Osterbrock's classification \citep[][]{Osterbrock1989} it corresponds to a type 1.8 AGN. The H${\alpha}$ profile shows an extreme, broad double-peaked line. Figure \ref{broad613} shows the multicomponent spectral fitting done in the H${\alpha}$ region. In particular, 
%we clearly see 
the presence of a broad component in H${\alpha}$ is clear, as well as a double-peaked H${\alpha}$ with two additional broad components, 
%\LEt{Alternatively, "the presence...components, is clear".} 
i.e., one component to the blue and the other to the red of the central line. In addition, we have fitted Gaussian functions to the narrow components of H${\alpha}$, [\ion{O}{i}]6300,6364, [\ion{N}{ii}]$6548,6584,$ and the [\ion{S}{ii}] doublet. This double-peaked profile observed in UNAM-KIAS 613 is similar to that present in a small fraction of AGNs \citep[less than 3\%;][]{Strateva+2003} as is the case in Arp 102B \citep{Popovic+2014} and in other double-peaked line emitters detected in the SDSS \citep{Wang+2005}. The full width half maximum (FWHM) of the whole broad H${\alpha}$ profile of UNAM-KIAS 613 achieves  $\sim$ 16\,000 km s$^{-1}$ and the separation between the blue and red broad components is larger than 10\,000 km s$^{-1}$. Different models have been suggested as possible explanations of the broad double-peaked low ionization emission lines in AGNs, although the preferred interpretation for the measured parameters in the H${\alpha}$ of UNAM-KIAS 613 corresponds to emission from a relativistic accretion disk model \citep[e.g.,][]{Eracleous+2009,Gezari+2007}.
This galaxy is reported as Seyfert 1.9 in the compilation by \cite{Veron-Cetty2006}.

%%%FigA613

%%%FigAbroad

\subsection{UNAM-KIAS 1197}

\begin{figure*}
\centering
\includegraphics[width=15cm]{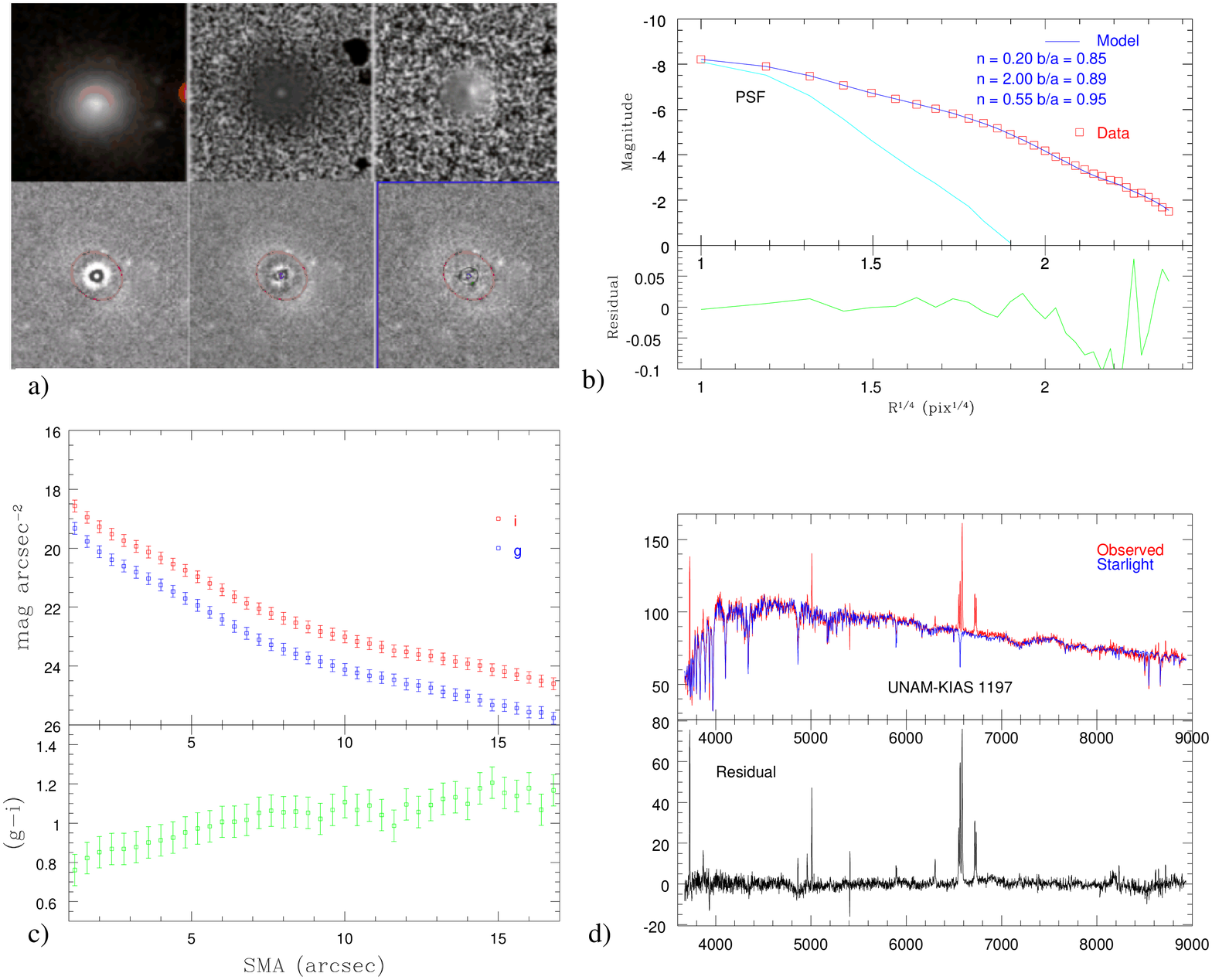}
\caption{
Similar to Fig. \ref{mosaico359}, but for UNAM-KIAS 1197.
}
\label{mosaico1197}
\end{figure*}

No similar-sized neighbor galaxy is found within %18 arcmin
692 kpc
and $\pm$1000 km s$^{-1}$.
Our images in the top-left  panel of Fig. \ref{mosaico1197} distinguishes a conspicuous bright central 
region with more uniform blue color along the face of this galaxy and an outer diffuse red region.

The 2D residual images show some central localized patches, probably associated with dust structures. 
The sloshing is not significant in the inner kpc. There is evidence of external low surface brightness shell-like structures.

Our analysis from GALFIT (top-right panel of Fig. \ref{mosaico1197}) shows three components: an inner component ($r_{eff}$ = 0.52 kpc) with S\'ersic 
index $n = 2.0$, another intermediate component ($r_{eff}$ = 2.05 kpc) with $n = 0.55$, and an outer component ($r_{eff}$ = 5.50 kpc) with $n = 0.2$.

At the distance of UNAM-KIAS 1197, a 3 arcsec fiber spectrum subtends 1.92 kpc. 
From our spectroscopic analysis, this galaxy qualifies as a LINER (see Fig. \ref{BPT}).
We note the intensity of \ion{N}{ii} with respect to H${\alpha}$ in the bottom-right panel of Fig. \ref{mosaico1197}. 
This feature is usually %identified
associated with Seyfert 2 nuclei. 
% IVAN: no entendi bien eso de Seyfert 2 si fue clasificado como LINER
However, the galaxy is reported as a Seyfert 1 with broad Balmer lines in the AGN
compilation by \cite{Veron-Cetty2006}.

%%%FigA1197

\subsection{UNAM-KIAS 1394}

\begin{figure*}
\centering
\includegraphics[width=15cm]{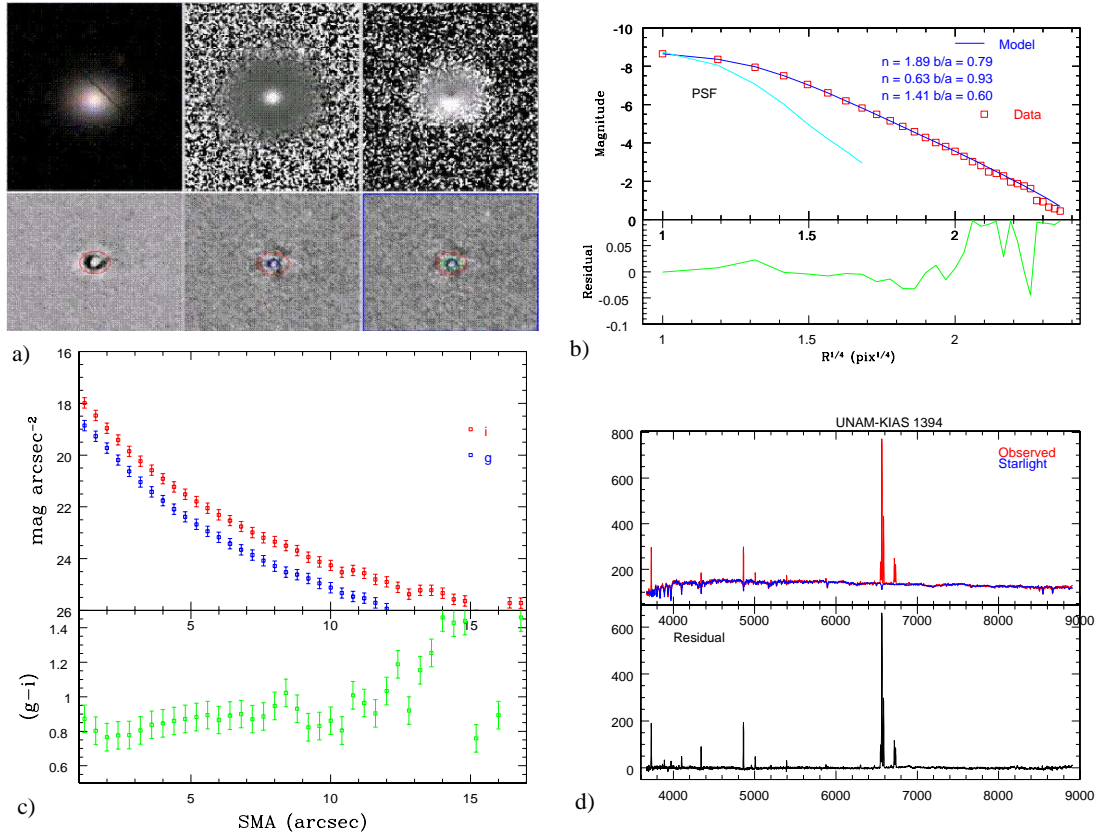}
\caption{Similar to Fig. \ref{mosaico359}, but for UNAM-KIAS 1394.}
\label{mosaico1394}
\end{figure*}

Only two neighbor galaxies fainter than 2.5 mag with respect to UNAM-KIAS 1394 are within %17 arcmin 
726 kpc
and $\pm$1000 km s$^{-1}$. 
The top-left panel of Fig. \ref{mosaico1394} shows a compact nuclear region and a diffuse outer region. The color 
map shows bluer colors along the face of this galaxy and a clear color gradient in the central region.

Our 2D residual images show a set of localized filaments in the central region, probably associated with dust structures. 
This galaxy does not show any significant sloshing in the inner kpc. We do not detect any low surface brightness features 
in this galaxy.

Top-right panel of Fig. \ref{mosaico1394} shows %three main components; 
two inner components ($r_{eff}$ = 0.60 and 
0.72 kpc) with S\'ersic indices $n = 0.63$ and 1.41, respectively, and a more extended intermediate component ($r_{eff}$ = 1.71 kpc) with 
$n = 1.89$. In this case, a solution with only one inner component in addition to the intermediate component is also reasonable.

At the distance of UNAM-KIAS 1394, the 3 arcsec fiber spectrum subtends 2.1 kpc.
The spectroscopic analysis shows definite star formation in the nuclear region (SFN) for this galaxy (see Fig. \ref{BPT}).

%%%FigA1394

%%% TABLA Phot
\begin{table*}
\caption{Photometric properties of the blue SF isolated elliptical galaxies. 
%classified as blue faint, red SF and red passive galaxies. 
}
  \centering
\begin{tabular}{c c c c c c}
\hline
\hline
name  &          fine structure    &      components    & $r_{eff}$  & $n$ \\ 
%(1)    &            (2)   &             (3)           & (4)         & (5)     \\
\hline
UNAM-KIAS 359 &   marginal LSB outer shells, &  inner & 0.15 &  0.10 \\
              &    dusty central region         &  inner & 0.55 &  0.15 \\                  
              &                              &  inner & 0.80 & 1.95 \\
              &                         &      intermed & 0.30 &   1.00 \\
UNAM-KIAS 613 &  dusty central region   &      inner &    0.43  &   1.47 \\
 &   & intermed & 2.81 &    0.55  \\
 &   &  outer & 4.54  & 6.50 \\            
UNAM-KIAS 1197&    LSB outer shells,                    & inner & 0.52 & 2.00\\  
              & dusty central region                    & intermed & 2.05 & 0.55 \\
        &                                               & outer & 5.50 & 0.20\\
UNAM-KIAS 1394 &      dusty central region  &         inner & 0.60 & 0.63 \\
   &    &               inner &     0.72 & 1.41    \\
   &    &       intermed  &  1.71  &         1.89 \\
\hline
\end{tabular}
\tablefoot{Columns are: name in the UNAM-KIAS catalog, type of fine structure features found after the 2D image decomposition, components after a 2D image decomposition along with their corresponding effective radii (kpc, $h = 0.73$), and S\'ersic indices. %respectively.  
}
\label{tabla_Phot}
\end{table*}

%%% TABLA Spec
\begin{table*}
\caption{Spectroscopic properties of the blue SF isolated elliptical galaxies. 
}
  \centering
\begin{tabular}{c c c c c c c c c}
\hline
\hline
  name  &    fiber scale  & nuclear spectrum & log(age$_{lw}$) & error  &  log(age$_{mw}$) & error & $\Delta$(age) \\
\hline
UNAM-KIAS 359  &   1.04   &         TO      &       9.00  &     1.32   &     9.98 &  0.32 & 8.5 \\ 
UNAM-KIAS 613  &     1.62 &   TO(AGN)   &       8.95  &  1.63      &    10.03 & 0.31  & 9.8 \\
UNAM-KIAS 1197 & 1.92     &   LINER         &       8.77  & 1.00       &    9.89  & 0.43  & 7.2 \\
UNAM-KIAS 1394 &     2.10 &          SFN    &        8.56 &  1.31      &    9.94  & 0.37  & 8.3 \\
\hline
\end{tabular}

\tablefoot{Columns are: name in the UNAM-KIAS catalog, physical scale (kpc, $h = 0.73$) subtended by the 3 arcsec SDSS fiber at the distance of each galaxy, type of spectrum and ionization source associated with the nuclear region from the modeled SDSS spectrum, log$_{10}$ of the average light-weighted age (in units of yr) %\textbf{$h = 0.73$?})
and its error, %Column (6)-(7): 
log$_{10}$ of the average mass-weighted age (in units of yr) %, \textbf{$h = 0.73$?}) 
and its error, and the star formation timescale (in units of Gyr).
}
\label{tabla_Spec}
\end{table*}

\section{Methodology of the photometric analysis} 
\label{Ap_Phot}

The homogeneous collection of 0.39 arcsec/pixel optical images available in the SDSS 
was used to carry out a photometric characterization of our E galaxies, in particular
the four blue SF isolated  galaxies. 
%Table \ref{tabla_Phot} presents a summary of the main photometric results obtained for these galaxies.
Searching for fine structure in the inner/outer regions of elliptical galaxies  
can be achieved with filter enhancement of the images. This procedure involves
taking the original image of the galaxy and filtering it with a Gaussian 
kernel of two/three times the size of the features  to enhance.    
Filter-enhancement was applied to the $r$-band SDSS images. Features in elliptical galaxies can also be 
revealed by color maps. Since we possess data from the SDDS $ugri$ bands, we generate $g-i$ color maps 
to look for color features/gradients toward the center.  

As the most important procedure in the present %study
section we choose the method of modeling the two-dimensional 
(2D) light distribution through GALFIT \citep{GALFIT2002, GALFIT2010}. %(Peng et al. 2002;2010). 
From the 2D image decomposition we try to reproduce the observed surface 
brightness distribution and explore how these local isolated elliptical galaxies may contain photometrically distinct
substructures that can shed light on their evolutionary history.

GALFIT can fit an arbitrary number or combination of parametric functions. For the present study, the S\'ersic 
function is adopted because (i) it is appropriate enough to model the main structural components in galaxies and because 
(ii) it has been widely used in the literature, thus it is useful for further comparisons with published results.
Each galaxy is fit with a series of models, each consisting of one to four S\'ersic components. 
These components were ranked by physical size (according to their effective radius, $r_{eff}$) and were generically designated 
as inner, intermediate and outer components. As a first order 
check of our multicomponent fitting, we extracted the azimuthally averaged 1D surface brightness profile from our best 2D 
models and compared it with the corresponding profile derived from the original data.

PSF images were built
by following procedures in the IRAF\footnote{http://iraf.noao.edu/} DAOPHOT package. Good estimates of seeing profiles are mandatory since 
residual errors in the seeing estimate could lead to mismatches between the model and the data. The different components 
are always fit by sharing the same central position. Since the presence of field stars can introduce potential uncertainty 
into the GALFIT component models, these stars were PSF subtracted or interpolated either close to the galaxy center or 
in the neighborhood previous to any fitting. At the end of the modeling, we carefully inspect the residual image with the original overlaid.

As a by-product of our image analysis, a measure of the nonaxisymmetry in the surface brightness distribution 
in the inner kpc of the UNAM-KIAS isolated elliptical galaxies is presented. The $r$-band images from the SDSS database  
were analyzed by fitting elliptical isophotes whose centers were first (i) kept fix and then (ii) allowed to vary. 
A comparison of the centers of the isophotes in (i) and (ii) may reveal a sloshing pattern or spatial variation in the 
central kpc that could indicate mass asymmetry and/or a dynamically unrelaxed behavior in the central regions of these 
galaxies. 

Surface brightness profiles in $g$ and $i$ bands were extracted by imposing a fixed center and also fixed ellipticity 
$\epsilon$ and position angle estimates during  our isophotal analysis. This guarantees homogeneous color estimates. The $g-i$ 
color gradient is corrected for mean galactic extinction.

\subsection{Residual features and fine structure} 
\label{fine-structure}

One of the most direct ways to study the assembly rate of elliptical galaxies is to measure the incidence of fine 
structure features. In this study, we examined the residual images after a 2D decomposition using additional 
enhancement procedures. This helped us (i) to identify problems related to poor modeling and also (ii) to identify 
possible stellar morphology disturbances in various forms. From the latter, we point out \emph{shells} that have been observed in 
the stellar component of nearby ellipticals, considered as the result from an accretion of a small companion by a 
massive galaxy \citep[e.g.,][]{DuprazCombes1986,Cooper+2011}. These shells  vanish after a few Gyr \citep{Schauer+2014}, 
so  they are indicators of a recent interaction. 
\emph{Tidal tails} are also recognized as linear streams of stellar matter, which is  evidence of 
a dynamically cold component in an accreted companion. Tails that are produced in this manner appear wide and short-lived 
in comparison to the long and narrow tails produced by interacting spirals. In addition, we note \emph{broad fans of stellar light} typically of low 
surface brightness, which are hard to detect in shallow surveys like the SDSS and, finally, \emph{highly disturbed galaxies} 
showing signs of an ongoing merger that may be disturbing the stellar component.

Although the more recent multiwavelength surveys suggest that the presence of dust in E/S0 galaxies is the rule rather than 
the exception, we note here that (i) the distribution of features found in our residual images after a 2D  decomposition 
were presumably associated with dust and that (ii) they are found concentrated on the nucleus and extending out to radii of 
a few hundred parsecs, rather than spread throughout the galaxy as in other samples  \citep{Finkelman+2010}.

\section{Methodology of the spectroscopic analysis}
\label{Ap_Spec}
%%%

We use the SDSS spectrum of each blue SF isolated elliptical galaxy to infer some stellar population and ionized gas properties.  
The retrieved SDSS spectra were obtained through an optical fiber of 3 arcsec. Those spectra  
include not only the nuclear emission but also the absorption bands and 
continuum produced by the stellar populations of the host galaxy. Thus, it  is necessary to subtract 
the stellar population contribution. This effect is especially important in galaxies of earlier morphological types 
where the observed spectrum is dominated by the stellar continuum, mainly of G and K-type giant stars, and by 
middle and old age stellar populations. For this purpose, we have applied 
%We have additionally applied 
STARLIGHT %\citep{STARLIGHT+2005} 
models to the observed SDSS spectra.
The models are based on 
a combination of 45 simple stellar populations (SSPs) from \citet[][]{BC2003}
with a Chabrier IMF. The SSP library used here consists of three metallicities 
Z (0.004, 0.02 and 0.05) for 15 ages between 0.001 and 13.0 Gyr. The intrinsic extinction
was modeled as a uniform dust screen, adopting the extinction law by \cite{Cardelli+1989}. 
%the method of principal 
%components analysis (PCA) by using eight eigen-spectra determined by  \cite{Hao2005}
%with non-emission spectra of SDSS galaxies in the low redshift range. To account for the 
%presence of an intermediate age star population, an eigen-spectrum of an A type star was included, as well as   
%a tenth eigen-spectrum corresponding to a power law  to take into account the potential presence of 
%a non-thermal emission source.  
Spectral fits were carried out in the spectral region
between 3000 \AA{} and 9100 \AA.

After removing the underlying continuum of the galaxy, all resulting spectra were reviewed to identify the presence of 
emission lines to construct the BPT diagram. In the spectral range covered by the SDSS spectra, the main detected emission 
lines correspond to H${\beta}$ 4863, [\ion{O}{iii}]5007, [\ion{O}{i}]6300, doublet [\ion{N}{ii}] 6548,6584, H${\alpha}$ 6583,{} and sulfur doublet [\ion{S}{ii}]6717,6731. The measurement of these lines was carried out 
interactively by Gaussian fitting with the SPLOT task in IRAF.  We also use NGAUSSFIT routine when more than one Gaussian was 
needed to fit the shape of an emission line, for example, because of the presence of broad line components.

The spectroscopic properties of each galaxy are %summarized in Table \ref{tabla_Spec} and presented
shown in the bottom-right panels of Figs. \ref{mosaico359}, \ref{mosaico613}, \ref{mosaico1197}, and \ref{mosaico1394}, showing simultaneously (i) the observed 
spectrum, (ii) the best fit to the observed  spectrum from the %
STARLIGHT model, and, finally, (iii) the residual spectrum 
after subtracting the observed and best-fit spectrum.

\begin{figure}
\includegraphics[width=8.8cm]{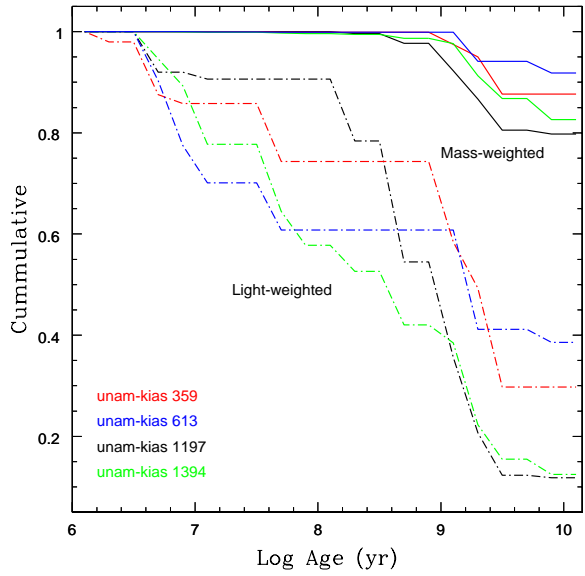}
\caption{Cumulative stellar population synthesis distributions %with STARLIGHT 
as a function of the
lookback time for
light-weighted ages %($h=0.73$??)
(dot-dashed lines)
and mass-weighted ages (solid lines) of the blue SF isolated E galaxies 
(UNAM-KIAS 359 in red, UNAM-KIAS 613 in blue, UNAM-KIAS 1197 in black, and UNAM-KIAS 1394 in green). 
%Color lines correspond to the color labels of the UNAM-KIAS galaxies.
}
\label{age_STAR}
\end{figure}

We  also applied the STARLIGHT population synthesis models to infer the star formation history (SFH) of these galaxies.
The results are presented in Fig. \ref{age_STAR} 
%\textbf{En caso que vaya esta figura, hacerla solo para las 4 blue SF aisladas. SEGUIR}
as the corresponding cumulative distributions of light-weighted and mass-weighted ages 
of the stellar populations that better reproduce the observed spectra. 
The average values are reported in Table \ref{tabla_Spec}.

\end{appendix}
\end{document}